\documentclass[superscriptaddress,aps,twocolumn,longbibliography,prl]{revtex4-1}
\usepackage{amsmath,amssymb}
\usepackage{graphicx}
\usepackage{braket}
\usepackage{color}
\usepackage{epsfig}
\usepackage{epstopdf}
\usepackage{hyperref}
\usepackage{bibentry}
\hypersetup{
colorlinks = true,
linkcolor = [rgb]{0.7,0.,0.},
citecolor = [rgb]{0.25, 0.41, 0.88},
urlcolor = [rgb]{0.25, 0.41, 0.88}}

\graphicspath{{figures/}}

\begin{document}
\title{ Tensor network wavefunction of $S=1$ Kitaev spin liquids }

\author{Hyun-Yong Lee}
\affiliation{Institute for Solid State Physics, University of Tokyo, Kashiwa, Chiba 277-8581, Japan}

\author{Naoki Kawashima}
\email{kawashima@issp.u-tokyo.ac.jp}
\affiliation{Institute for Solid State Physics, University of Tokyo, Kashiwa, Chiba 277-8581, Japan}

\author{Yong Baek Kim}
\email{ybkim@physics.utoronto.ca}
\affiliation{Department of Physics, University of Toronto, Toronto, Ontario M5S 1A7, Canada}

\date{\today}

\begin{abstract}
	Recently there has been a great interest in understanding quantum spin liquid phases with varying spin magnitude, partly due to possible material realizations. A number of recent numerical computations suggest that the ground state of the S=1 Kitaev model may be a quantum spin liquid in analogy to the renowned $S$=$1/2$ model. On the other hand, the nature of the ground state remains elusive since the $S$=$1$ model is not exactly solvable in contrast to the $S$=$1/2$ model. In this work, we construct a tensor network ground state wavefunction for the S=1 Kitaev model, which is explicitly written in terms of physical spin operators. We explain how this class of wavefunctions can be successfully used for variational computations and compare the outcomes to known results on finite size systems. We establish the existence of distinct topological sectors on torus by constructing the minimally entangled states in the degenerate ground state manifold and evaluating topological entanglement entropy. Our results suggest that the ground state of the S=1 Kitaev model is a gapped quantum spin liquid with Z2 gauge structure and Abelian quasiparticles.
\end{abstract}
\maketitle

{\it Introduction - } The exact solution of the $S$=$1/2$ Kitaev model on the honeycomb lattice, i.e., Kitaev spin liquid\,(KSL), is an important milestone in theoretical research on quantum spin liquids\cite{Kitaev2006}. It has firmly established the existence of quantum spin liquids in systems with local spin interactions and opened the door to possible quantum spin liquid states arising from the bond-dependent competing interactions. Both aspects lead to recent intensive experimental and theoretical explorations of the Kitaev materials\cite{Khaliullin2005, Jackeli2009, Plumb2014, Sears15, Johnson15, Kim15, Kim16, Yadav16, Zhou2016b, Banerjee2016, Luke16, Sinn16, Winter16, Leahy17a, Trebst2017, Banerjee17, Catuneanu18, Gohlke18, Winter18, Banerjee2018, Balz19, Wang19}, where the Kitaev interaction is dominant 
and the bond-dependent interactions originate from strong spin-orbit coupling. 

On the other hand, the nature may also allow the Kitaev's bond-dependent interactions for higher-spin magnitudes, as recently suggested by electronic structure computations\cite{Peter19}. While the large-spin limit would simply correspond to the classical limit, the ground state of the Kitaev model for relatively small spin magnitudes, especially for S=1, has not been fully understood as the higher-spin Kitaev models on honeycomb lattice do not have exact solutions. It was pointed out early on that there exists a flux operator, $W_p$, made of six spin operators on each plaquette for an arbitrary spin $S$, which commutes with the Hamiltonian\cite{Baskaran08}. Hence the ground state must have a definite eigenvalue of the flux operator, just like the case of the $S$=$1/2$ model. Exact diagonalization of the $S$=$1$ Kitaev model on small system sizes show that the ground state is non-magnetic\cite{Koga18}. Thermal pure quantum state approach at finite temperatures on small system sizes finds an entropy plateau, similar to the $S$=$1/2$ model\cite{Oitmaa18}. All of these indicate the possibility of a quantum spin liquid in the $S$=$1$ Kitaev model while the precise nature of the ground state is not known.

In this paper, we provide the tensor network\,(TN) representation of the ground state wavefunction for the $S$=$1$ Kitaev model, with both ferromagnetic and antiferromagnetic interactions between the $S$=$1$ local moments. Our approach is inspired by the recent construction of the TN ground-state wavefunction of the $S$=$1/2$ Kitaev model\cite{HY19}, where the wavefunction is written in terms of physical spin variables, instead of the Majorana fermion operators in the exact solution. We start from the bond dimension $D$=$2$ TN representation of the wavefunction, which is an explicit eigenstate of the flux operator via the projector $Q_{\rm LG}$, or the Loop Gas\,(LG) operator. The resulting wavefunction can be regarded as the sum over loop gas configurations made of a series of local actions of $Q_{\rm LG}$ along a loop of lattice sites. We show that the $Z_2$ gauge structure is automatically built in the wavefunction. Using the mapping between the norm of the wavefunction and the partition function of the classical loop gas, it is shown that the wavefunction is in a gapped state, which is corroborated by the direct computation of the correlation length from the transfer matrix. 
This wavefunction is further improved by applying an additional Dimer Gas\,(DG) operator\cite{HY19}
and imaginary time evolution\,(ITE), resulting in 
excellent variational energy.

Our approach also allows us to understand some partial information about topological properties of the ground state wavefunction on torus via the construction of minimally entangled states in the ground state manifold. Using this construction, we contrast the difference between the degenerate ground states of the $S$=$1$ and $S$=$1/2$ systems, elucidating the nature of the quantum spin liquid ground state in the $S$=$1$ Kitaev model.

\begin{figure}[!t]	
	\includegraphics[width=0.5\textwidth]{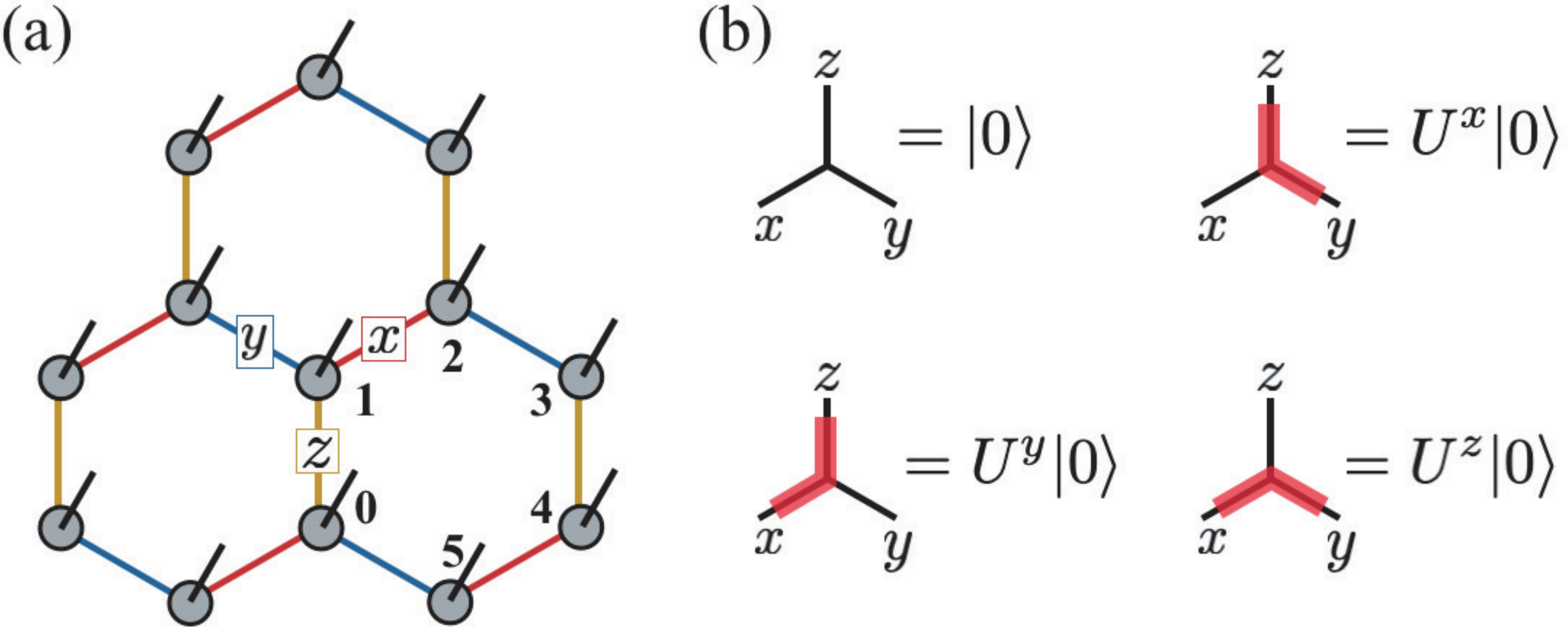}
	\caption{ (a) Graphical representation of the tensor network state on the honeycomb lattice where the $x,y$ and $z$ bonds of the model in Eq.\,\eqref{eq:hamiltonian} are denoted by red, blue and yellow colors, respectively. (b) The local states depending on the local loop configuration appearing in the LG state\,[Eq.\,\eqref{eq:lg_state}]. Here, $|0\rangle$ stands for the (111)-direction magnetic state, i.e. $\langle 0|S^\gamma |0\rangle$=$1/\sqrt{3}$, and $U^\gamma = e^{i\pi S^\gamma}$ is the SO(3) rotation operator. }
	\label{fig:schematic}
\end{figure}

{\it Model and symmetry - } Hamiltonian of the Kitaev model on the honeycomb lattice\cite{Kitaev2006} is defined as 
\begin{align}
	H_{ij}^\gamma = -K \sum_{\langle ij \rangle_\gamma} S_i^\gamma S_j^\gamma,
	\label{eq:hamiltonian}
\end{align}
where $\langle ij\rangle_\gamma$ stands for the nearest neighboring sites, $i$ and $j$, on the $\gamma$-bond with $\gamma = x,y,z$ as depicted in Fig.\,\ref{fig:schematic}\,(a), and $S^\gamma$ is the spin-1 operator. One can verify that the Hamiltonian commutes with a flux operator $W_p = U^x_0 U^y_1 U^z_2 U^x_3 U^y_4 U^z_5 = \pm 1$ where $U^\gamma = e^{i \pi S^\gamma}$ is the $180^\circ$ spin-rotation operator along the $\gamma$-direction, and sites 0-5 are defined in Fig.\,\ref{fig:schematic}\,(a). It indicates that the spin-1 model also exhibits the $Z_2$ gauge redundancy, and thus the Hilbert space can be sectorized by the combination of the flux number.

{\it Loop gas states. -} In a similar fashion to the spin-$1/2$ model\cite{HY19}, one can define the LG operator $Q_{\rm LG}$ for the spin-1 model in a bond dimension $D$=$2$ TN representation with a local tensor
%
	$Q_{\lambda \mu \nu} = \tau_{\lambda \mu \nu} (U^x)^{1-\lambda} (U^y)^{1-\mu} (U^z)^{1-\nu}$,
%
where $\lambda,\mu,\nu = 0,1$ are the virtual indices. Here, non-zero elements of the tensor $\tau$ are only $\tau_{\lambda \mu \nu} = -i$ for $\lambda$=$\mu$=$\nu$=$0$ and $1$ for $\lambda$+$\mu$+$\nu$=$2$. It will be shown later that this operator generates loop configurations made of local actions represented in Fig.\,\ref{fig:schematic}\,(b). One can examine the physical symmetries of $Q_{\rm LG}$ at the local tensor level as done in Ref.\,\cite{HY19} and also derive the $Z_2$ gauge symmetry: $\sum_{\lambda',\mu',\nu'} \sigma^z_{\lambda \lambda'} \sigma^z_{\mu \mu'} \sigma^z_{\nu \nu'} Q_{\lambda' \mu' \nu'} =  Q_{\lambda \mu \nu}$. Furthermore, it is straightforward to show that the local $Q$-tensor obeys the following equations
\begin{align}
	U^x Q_{\lambda \mu \nu} = \sum_{\mu',\nu'} \sigma^x_{\mu\mu'}\sigma^x_{\nu\nu'} Q_{\lambda \mu' \nu'}, \nonumber\\
	U^y Q_{\lambda \mu \nu} = \sum_{\nu',\lambda'} \sigma^x_{\nu\nu'}\sigma^x_{\lambda \lambda'} Q_{\lambda' \mu \nu'}, \nonumber\\
	U^z Q_{\lambda \mu \nu} = \sum_{\lambda',\mu'} \sigma^x_{\lambda\lambda'}\sigma^x_{\mu\mu'} Q_{\lambda' \mu' \nu}, 
	\label{eq:key_relation_index}
\end{align}
where $\sigma^x$ is the Pauli matrix. Note that the $Q$-tensor for the spin-1/2 model obeys similar relations, but with a non-Hermitian unitary matrix $v$, defined in Ref.\,\cite{HY19}, rather than $\sigma^x$ in the above relations. That is because $U^\gamma$ satisfies the commutator relation, $[U^\gamma, U^{\gamma'}] = \delta_{\gamma \gamma'}$, while the Pauli matrices follow the anti-commutator relation, $\{\sigma^\gamma, \sigma^{\gamma'}\} = 2\delta_{\gamma \gamma'}$. In addition, as will be seen below, it brings about fundamental distinction between the $S$=$1/2$ and $S$=$1$ LG states including topological property, even though the difference\,($v \leftrightarrow \sigma^x$) may not look critical at first sight. One can also show at the local TN level that $Q_{\rm LG}$ projects any quantum state into the vortex-free sector using Eq.\,\eqref{eq:key_relation_index} or $W_p Q_{\rm LG} = Q_{\rm LG}$. See supplemental material\,(SM) for more details\cite{SM}. 

Due to the symmetries at the isotropic point, we apply the LG operator onto a classical product state $|0\rangle \equiv \otimes_i |0\rangle_i$ where all spins align in the $(111)$-direction, i.e., $\langle 0 | S_i^\gamma |0\rangle = 1/\sqrt{3}$. It results in the LG state $|\psi_{\rm LG}\rangle \equiv Q_{\rm LG} |0\rangle$,
\begin{align}
	\includegraphics[width=0.26\textwidth]{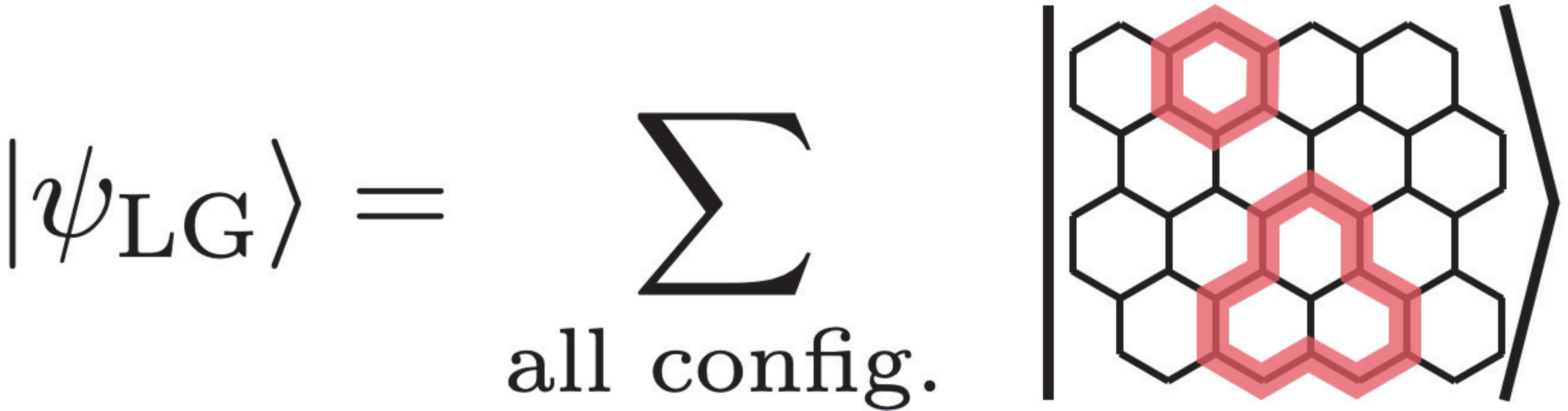}.
	\label{eq:lg_state}
\end{align}
This is, the equal weight superposition of all possible loop configurations where the local state depends on the direction of the local loop as depicted in Fig.\,\ref{fig:schematic}\,(b). Using the identity $\langle 0 | U^\gamma | 0 \rangle$=$1/3$, one can map the norm of $|\psi_{\rm LG}\rangle$ into the partition function of the classical $O(1)$ LG model with the fugacity $\zeta$=$1/3$ which is in the gapped phase of the model\cite{Nienhuis1982}. It is worth noting that the $S=1/2$ LG maps to the critical point $\zeta_c = 1/\sqrt{3}$\cite{HY19}. 

Utilizing the corner transfer matrix renormalization group\,(CTMRG), we have computed the energy expectation value\,(per site) of $|\psi_{\rm LG}\rangle$, $E_{\rm LG} = -0.50562$, which is far from the previous ones $E^{\rm 24site}=-0.648$ obtained on a 24-site system\cite{Koga19} and $E^{\rm DMRG} = -0.644$ obtained by the density matrix renormalization group method on a $(48\times3)$-cylinder system\cite{YB19}. In what follows, we present two different ways to vary the LG state to obtain better variational energies while most physical properties remain intact: (i) applying the so-called dimer gas\,(DG) operator onto the LG state and optimize its variational parameters as proposed in Ref.\,\cite{HY19}, (ii) evolving the LG state through the imaginary time evolution\,(ITE) operator\cite{Tao08}.

\begin{figure}[!t]
	\includegraphics[width=0.5\textwidth]{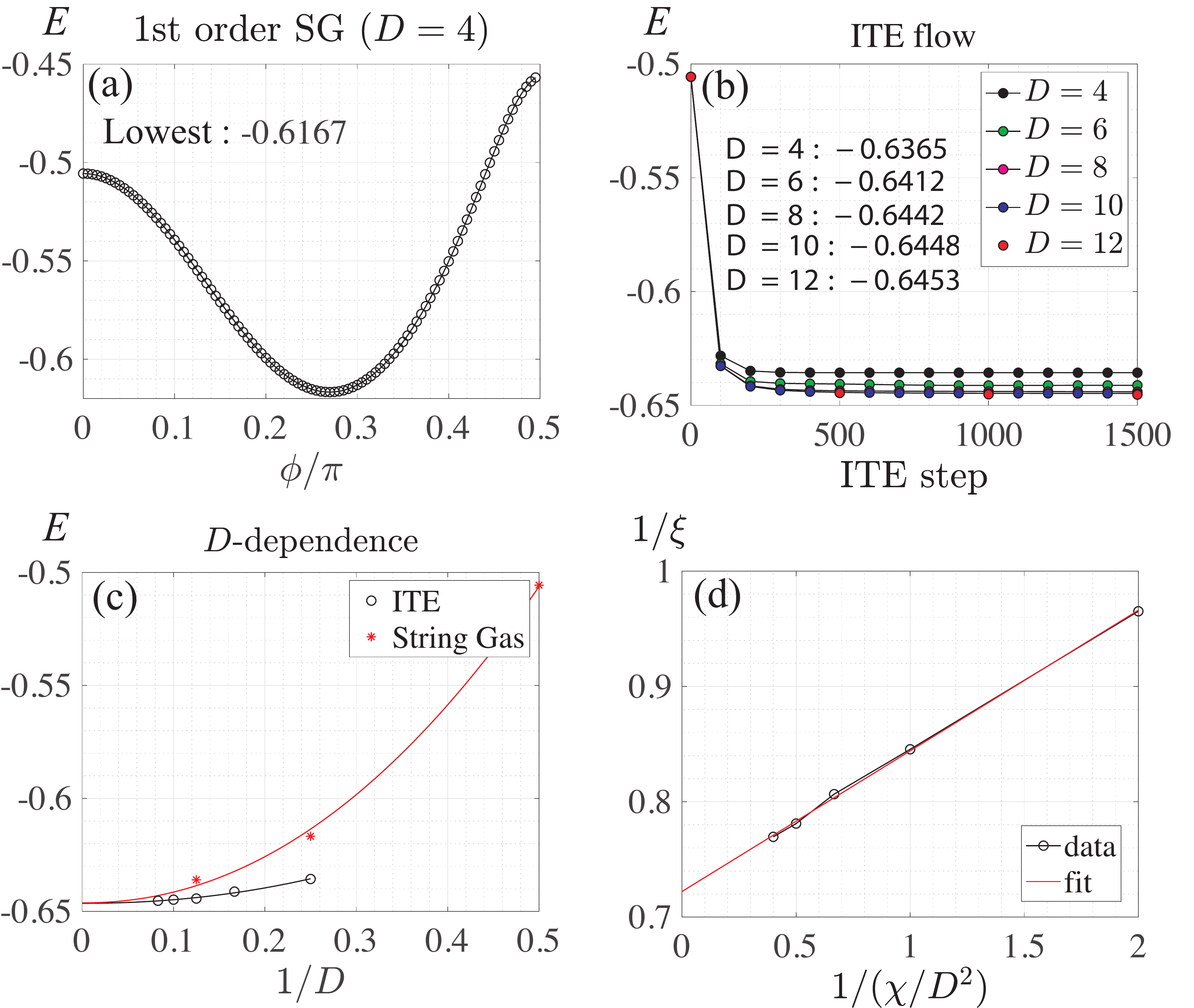}
	\caption{ Variational energies of (a) the first order SG state $|\psi_{{\rm SG}_1}(\phi)\rangle$ and (b) the ITE from the initial LG state at a given bond dimension $D$. (c) $D$-interpolation of the variational energies obtained from both SG ansatz and ITE method. (d) The largest correlation length $\xi$ of the $D$=$10$ ITE ansatz as a function of the CTM dimension $\chi$. }
	\label{fig:optimization}
\end{figure}

{\it String gas states. -} The DG operator $R_{\rm DG}$ is also represented in a $D=2$ TN with the following local tensor
%
	$R_{\lambda \mu \nu}(\phi) = \zeta_{\lambda \mu \nu}^\phi (S^x)^{1-\lambda} (S^y)^{1-\mu} (S^z)^{1-\nu}$,	
%
where the non-zero elements of $\zeta$-tensor are simply $\zeta_{000}^\phi = \cos\phi$ and $\zeta_{100}^\phi = \zeta_{010}^\phi = \zeta_{001}^\phi = \sin\phi$. 
Applying the DG operators onto the LG state results in the string gas\,(SG) state with variational parameters $\{ \phi_\alpha \}$, i.e., $|\psi_{{\rm SG}_n}(\{ \phi_\alpha \})\rangle \equiv \prod_{\alpha=1}^n R_{\rm DG}(\phi_\alpha) |\psi_{\rm LG}\rangle$, which can be optimized to have the lowest energy\cite{HY19}. Here, $n$ is referred to as the order of the SG state, and the bond dimension grows as $D=2^{n+1}$. The variational energy of the first order ansatz\,($D$=$4$) is shown in Fig.\,\ref{fig:optimization}\,(a) as a function of the variational parameter $\phi$. The lowest energy is found to be $E_{{\rm SG}_1}$=$-0.6167$ which is still significantly higher than the ones obtained by ED and DMRG. Note that the efficiency of the DG operator is not as good as the case of the $S$=$1/2$ model. We believe that this is because the $S$=$1$ DG operator cannot be written as the polynomial function of the Hamiltonian in contrast to the spin-$1/2$ model, where the first order SG state already gives $99.8\%$ accuracy to the exact one energetically\cite{HY19}. We have also optimized the second order ansatz $|\psi_{{\rm SG}_2}(\phi_1,\phi_2)\rangle$ and obtained the best energy $E_{{\rm SG}_2}$=$-0.6366$\,[see the SM\cite{SM} for more details]. 

{\it Imaginary time evolution. -} We perform the ITE by applying the two-site gate $e^{\tau K S_i^\gamma S_j^\gamma}$ on every bond and updating the site tensors using the simple update\,(SU) iteratively\cite{Tao08}. 
Note that it is almost impossible to find a KSL-like spin liquid, e.g., $W_p$=$1$, with SU from a random initial state. To be more precise, the resulting states are mostly magnetically ordered and not vortex-free, $\langle W_p \rangle$$<$$1$, and even exhibit the strong initial state dependence.
However, the initial LG state with a careful choice of $D$\cite{Mei17} makes the ITE quite stable and reliable. For instance, the ITE operator commutes with the LG operator, i.e., $[e^{\tau K S_i^\gamma S_j^\gamma}, Q_{\rm LG}]=0$, so that the ITE does not spoil the vortex-freeness. While the truncated singular value decomposition in the SU may destroy the vortex-free condition, it can be avoided by keeping the degenerate singular values. Figure\,\ref{fig:optimization}\,(b) presents the energy flow as a function of the ITE step. The zeroth step denotes the LG state, and the energy drops quickly and converges already in a few hundred steps. Throughout the ITE, the flux expectation value keeps unity, $\langle W_p\rangle$=$1$, while the state remains non-magnetic, $\langle \vec{S} \rangle$=$0$ up to the machine precision\,[see SM\cite{SM}]. The variational energy decreases monotonically with increasing $D$, and the lowest energy obtained by the ITE is $E_{\rm ITE}^{D=12}$=$-0.6453$ which is comparable to the ones predicted by ED and DMRG. Comparing to the energies of the SG states, the ITE energies are lower at the same $D$. However, this is not so surprising since the tensors of the SG states are quite sparse, i.e., many of elements are zero. To estimate the infinite-$D$ variational energy, we have extrapolated the energies $E_{\rm SG}$'s and $E_{\rm ITE}$'s, and both seem to converge to almost the same one $E_{D\rightarrow\infty}$$\simeq$$-0.6464$ as shown in Fig.\,\ref{fig:optimization}\,(c). Finally, to see how the gapped nature of the LG state is affected by the ITE, we have directly computed the largest correlation length $\xi$ from the largest and and second largest eigenvalues, $\lambda_0$ and $\lambda_1$, of the transfer matrix in the CTMRG algorithm: $\xi = -1/\log(\lambda_1/\lambda_0)$. The result from the $D$=$10$ ansatz is presented in Fig.\,\ref{fig:optimization}\,(d) as a function of the CTM dimension $\chi$. It clearly indicates a finite and short correlation length in contrast to the case of the spin-$1/2$ where $\xi$ diverges as increasing $\chi$\cite{HY19}. Additionally, the smooth connection from the SG ansatz to the LG state, which is strictly gapped, is a strong evidence of the gapped nature of the $S$=$1$ KSL. 

\begin{figure}[!t]
	\includegraphics[width=0.5\textwidth]{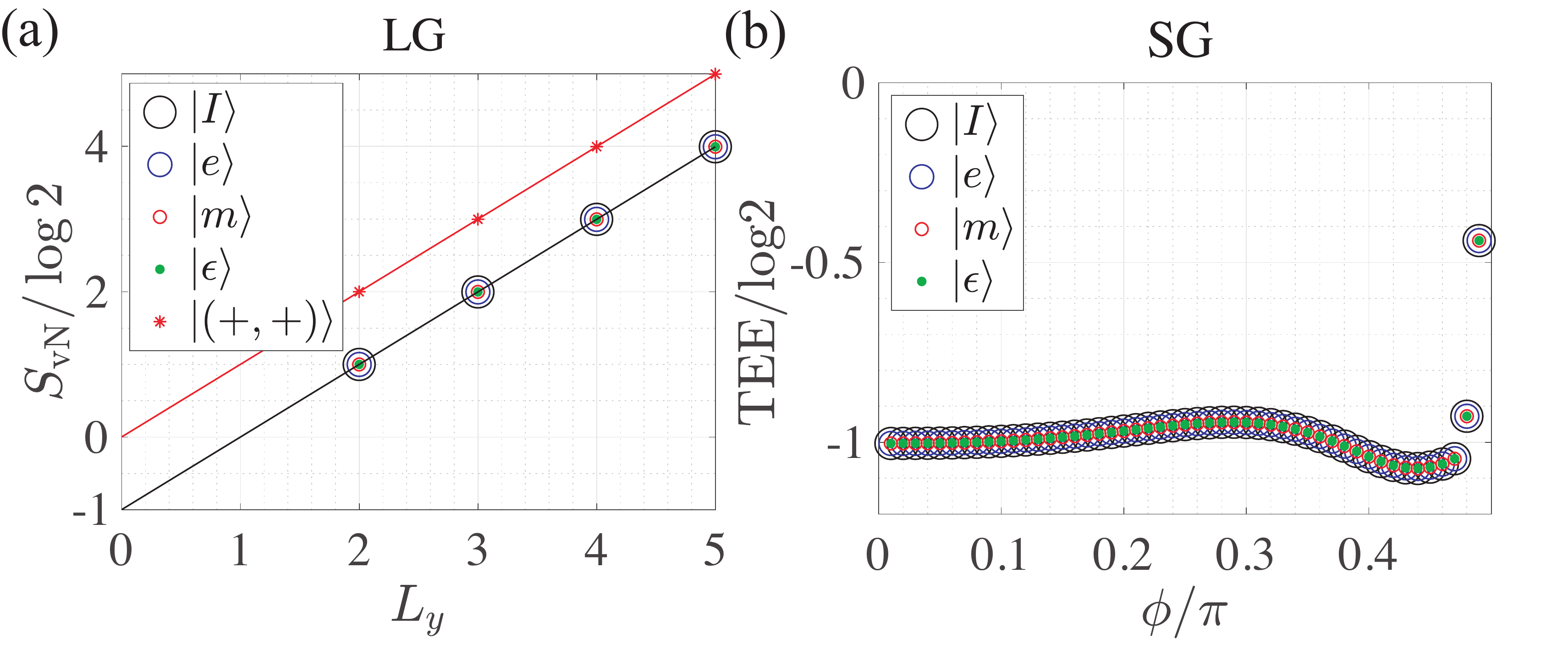}
	\caption{ (a) The entanglement entropy $S_{\rm vN}$ of the LG state as a function of the circumference $L_y$ in the $(+,+)$-flux and the MES sectors. (b) The extracted topological entanglement entropy of the first order SG state $|\psi_{{\rm SG}_1}(\phi)\rangle$. The definition and construction of four MESs, $|I\rangle$, $|e\rangle$, $|m\rangle$ and  $|\epsilon\rangle$ are presented in the SM\,\cite{SM}.}
	\label{fig:entropy}
\end{figure}

{\it Topological property. -} On a compact geometry, one should consider not only the local fluxes but also the global fluxes defined on non-contractable paths in the system. For instance, we are able to define a global flux operator on the cylinder, e.g., $W_\Gamma = \prod_{i\in \Gamma} U_i^y$ wrapping the cylinder. Then, one can sectorize the Hilbert space further using the global flux $W_\Gamma = \pm 1$. 
It has been found that, in the case of the spin-$1/2$, the sector is characterized by the parity of the number of non-contractible loops in the configuration\cite{HY19a}. For instance, the parity of the number of loops connecting two edges of the cylinder determines the sector. On the other hand, the $S$=$1$ LG state in each sector is characterized in a completely different way. Namely, all loop configurations are allowed regardless of the sector, whereas the sign of the configurations with the odd number of the loops winding the cylinder depends on the sector:
\begin{align}
	\includegraphics[width=0.45\textwidth]{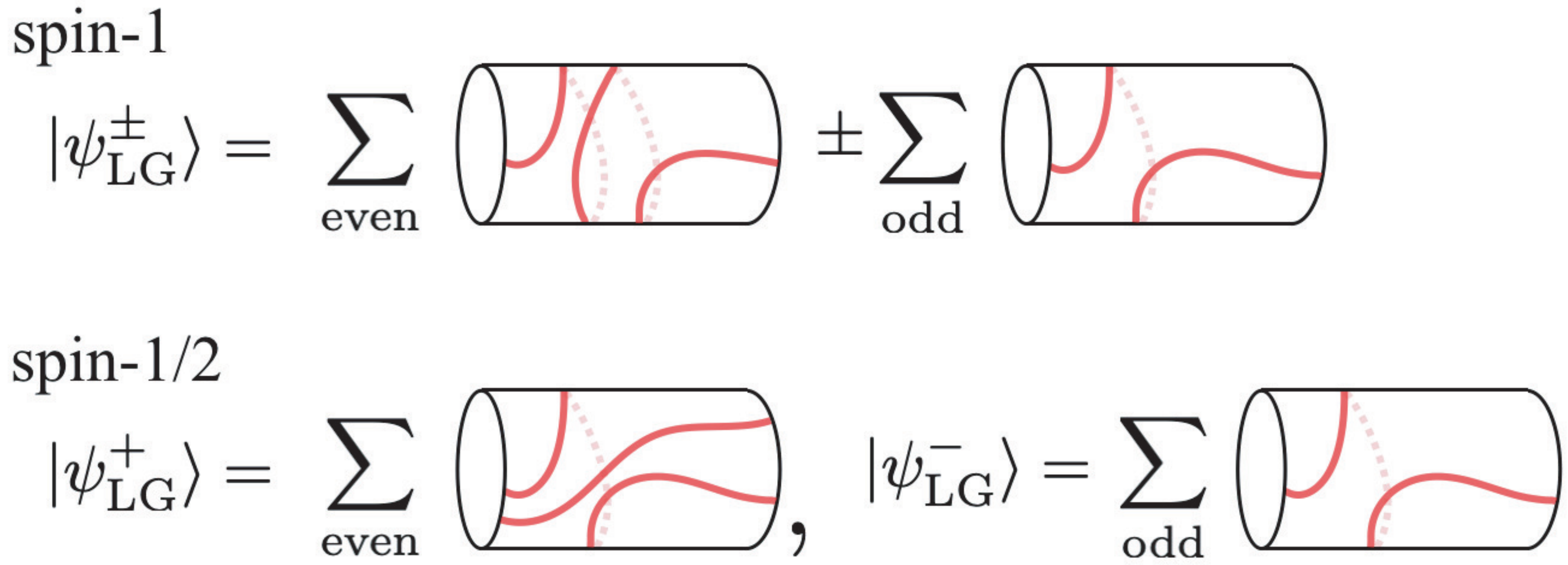}.\nonumber
\end{align}
This qualitative difference between the $S$=$1/2$ and $S$=$1$ LG states comes from Eq.\,\eqref{eq:key_relation_index} which is different from the $S$=$1/2$ case\,[see the SM\cite{SM} for more details]. It can be shown that the $(-)$-sector can be obtained by twisting the gauge of the LG operator using the non-trivial element of the $Z_2$ invariant gauge group\cite{SM}. Then, one can construct the minimally entangled states\,(MES) on the infinite cylinder\,(say, $L_x \rightarrow \infty$, $L_y:$  finite) using the degenerate states in the different flux sector\cite{Zhang12}. The explicit definition of the MES is presented in the SM\cite{SM}.  It is worth noting that one can construct all the MESs in the $S$=$1$ model in contrast to the $S$=$1/2$ model where only the trivial and vortex sectors are accessible\cite{HY19a}. Using the boundary theory of TN\cite{Cirac11}, we have calculated the entanglement entropy\,(EE) and then extracted the topological entanglement entropy\,(TEE) to identify the nature of each anyon sector. The result is shown in Fig.\,\ref{fig:entropy}\,(a), where the EE scales as $S_{\rm vN}=(\log 2)$$\times$$L_y-\gamma $, and TEE is $\gamma$=$0$ for $|(+,+)\rangle$ and $\gamma$=$\log 2$ for all other MESs. It indicates that all topological excitations are Abelian so that the ground state possesses the $Z_2$ topological order.
Figure\,\ref{fig:entropy}\,(b) presents the $\phi$-dependence of the TEE of $|\psi_{{\rm SG}_1}\rangle$. When the SG state is close to the LG state\,(near $\phi=0$), the TEE is almost perfectly $\log 2$. As $\phi$ increases, the TEE deviates from $\log 2$ which might come from the stronger finite size effect\,($L_y$). 



%
\begin{figure}[!t]
	\includegraphics[width=0.23\textwidth]{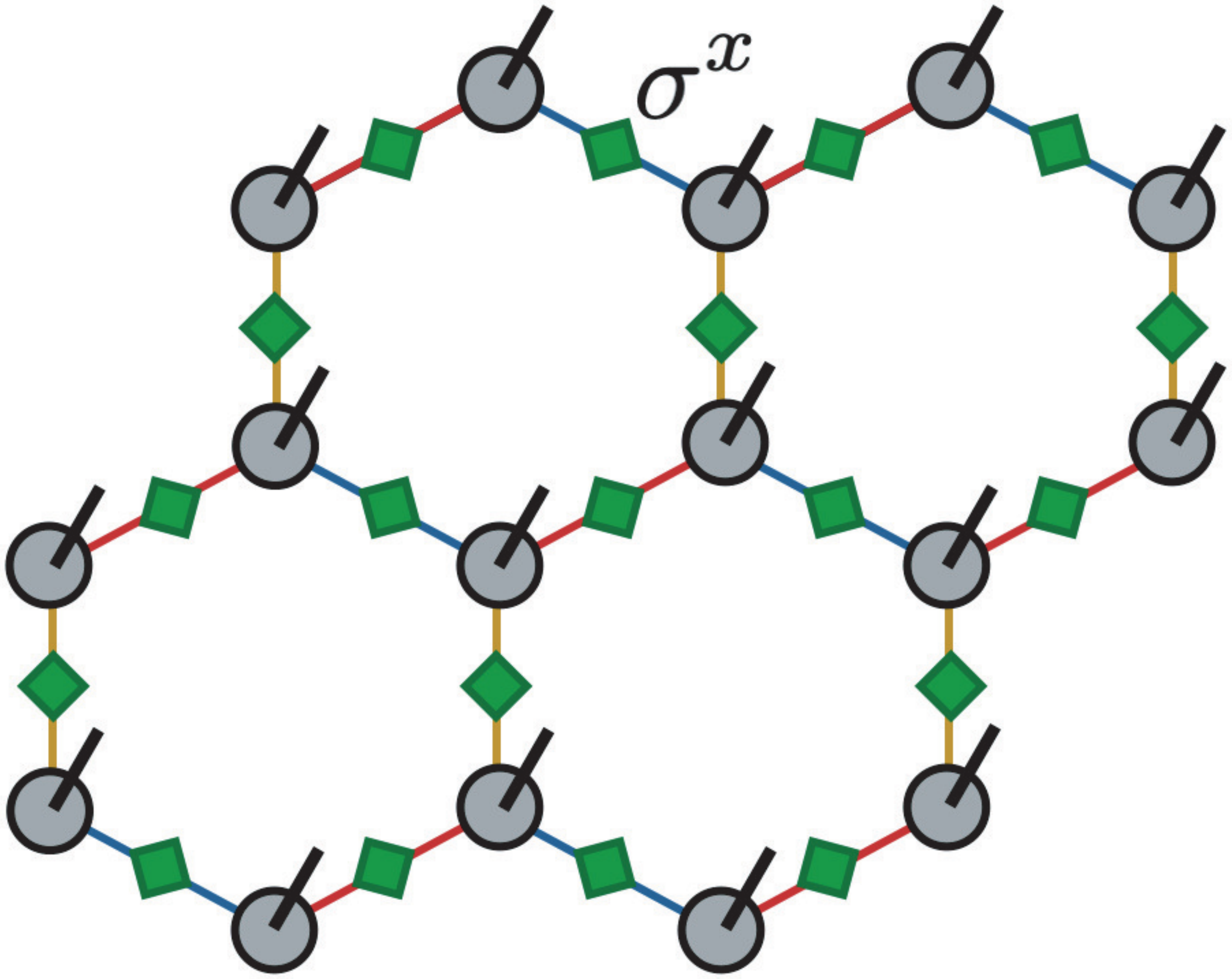}
	\caption{Schematic figure of the TN state for the antiferromagnetic model, where the green square denotes the Pauli $\sigma^x$-matrix, and the original TN state\,(without green squares) is either the LG state or the SG state.}
	\label{fig:afm_transform}
\end{figure}

{\it Antiferromagnetic model. -} Remarkably, the relation in Eq.\,\eqref{eq:key_relation_index} allows us to construct the ansatz for the antiferromagnetic model\,($K$$<$$0$) without much effort. First, we prepare a TN operator, say $\widetilde{Q}_{\rm LG}$, by inserting $\sigma^x$-matrix into every bond in the $Q_{\rm LG}$-operator as illustrated in Fig.\,\ref{fig:afm_transform}\,(a). Then, using Eq.\,\eqref{eq:key_relation_index}, one can show that the $\widetilde{Q}_{\rm LG}$-operator also guarantees the vortex-free condition, i.e., $W_p \widetilde{Q}_{\rm LG} = \widetilde{Q}_{\rm LG}$, and is related to the $Q_{\rm LG}$ by $\widetilde{Q}_{\rm LG} = V Q_{\rm LG}$ where $V$ is a unitary transformation flipping the overall sign of the Hamiltonian, i.e., $V^\dagger H V = -H$. Detailed derivation and proofs are provided in the SM\cite{SM}. It implies that the deformed LG state $|\widetilde{\psi}_{\rm LG}\rangle \equiv \widetilde{Q}_{\rm LG} |0\rangle$ gives exactly the same variational energy as that of the original LG state for the antiferromagnetic model: $\langle \widetilde{\psi}_{\rm LG}| H_{K<0} | \widetilde{\psi}_{\rm LG} \rangle$=$\langle\psi_{\rm LG}| H_{K>0} | \psi_{\rm LG} \rangle$=$E_{\rm LG}$. Besides, since the DG operator commutes with both $Q_{\rm LG}$ and $\widetilde{Q}_{\rm LG}$, one can also deform the SG state to have the same energy for the antiferromagnetic model. That is, the variational energy of $|\widetilde{\psi}_{{\rm SG}_n}(\{ \phi_\alpha \})\rangle = \prod_{\alpha=1}^n R_{\rm DG}(\phi_\alpha) |\widetilde{\psi}_{\rm LG}\rangle$ is simply $\langle \widetilde{\psi}_{{\rm SG}_n}(\{ \phi_\alpha \}) | H_{K<0} | \widetilde{\psi}_{{\rm SG}_n}(\{ \phi_\alpha \})\rangle = E_{{\rm SG}_n}(\{ \phi_\alpha \})$. Therefore, once the variational parameters are optimized for the ferromagnetic model, one can use the same ones for the antiferromagnetic model with insertion of the $\sigma^x$-matrix in the TN state. As will be shown below, this simple transformation comes in handy when we consider a weak perturbation, e.g., the magnetic field, in the antiferromagnetic model. It is also worth noting that such TN deformation for the $S$=$1/2$ model exists but requires a larger unit-cell in the TN state\cite{SM}.

\begin{figure}[!t]
	\includegraphics[width=0.5\textwidth]{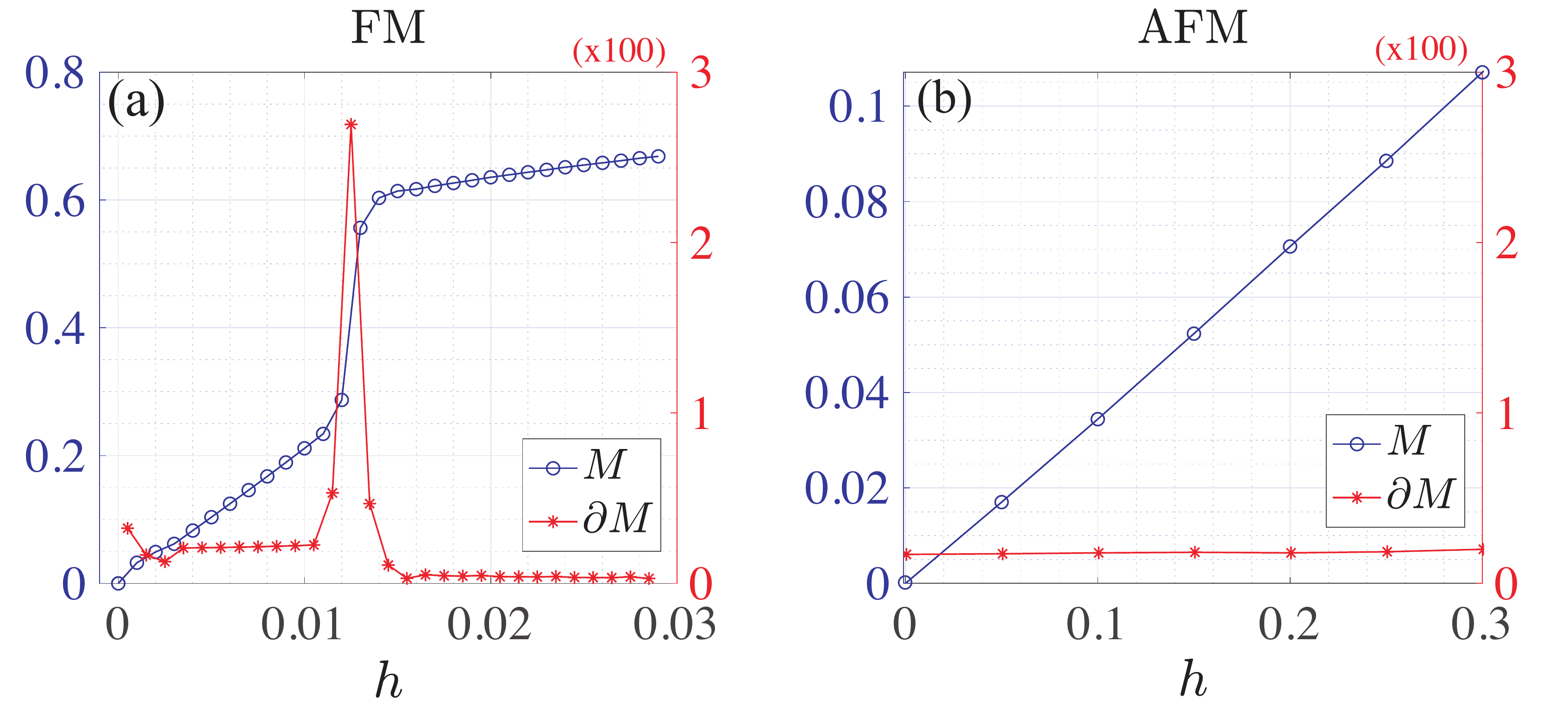}
	\caption{ The magnetizations of (a) the ferromagnetic model and (b) the antiferromagnetic model as a function of the field strength $h$. The results are obtained by the ITE with $D=12$. }
	\label{fig:field_result}
\end{figure}

{\it Magnetic field. -} Here, we discuss how the $S$=$1$ KSL is affected by the $(111)$-direction magnetic field, 
\begin{align}
	H_{\rm field} = - \frac{h}{\sqrt{3}}\sum_i (S_i^x+S_i^y+S_i^z).	
\end{align}
To this end, we perform the ITE from the LG and the deformed LG states for the ferromagnetic and antiferromagnetic models, respectively, in the presence of the magnetic field. We have found that the response to the field resembles the ones of the $S$=$1/2$ model\cite{Zhu18, Hickey19}. That is, the ferromagnetic KSL is easily driven to the polarized phase, in which the spins are aligned in the (111)-direction, at a weak field $h_c^{D=12} \approx 0.013 $ as shown in Fig.\,\ref{fig:field_result}\,(a), where the $D$=$12$ TN is optimized to determine the critical field. On the other hand, Figure\,\ref{fig:field_result}\,(b) indicates that the antiferromagnetic KSL is quite robust against the field where there is no transition at least below $h=0.3$. According to the second derivative of the energy and the first derivative of the magnetization and flux expectation value\cite{SM}, the transition occurring in the ferromagnetic model is expected to be continuous. It is also noteworthy that the $D$-dependence of the ansatz is stronger than that of the $S$=$1/2$ model\cite{HY19b}. The $D$-dependence of the results is discussed in the SM\,\cite{SM}.

{\it Conclusion. -} We have studied the ground state properties of the $S$=$1$ Kitaev model employing the LG, SG ansatz and the ITE optimization. The ground state is found to be a gapped $Z_2$ spin liquid which differs from the $S$=$1/2$ model hosting the gapless KSL ground state\cite{Kitaev2006}. The gapped nature has been shown by mapping the LG state into the partition function of the classical $O(1)$ LG model in the gapped phase and by directly computing the largest correlation length of the optimized ansatz. By constructing the MESs, we have identified the topological nature of the excitations, i.e., the Abelian anyons. The TN representation allows us to understand the qualitative difference between the $S$=$1$ and $S$=$1/2$ LG and SG states including the topological property. In addition, we have found a simple transformation between the tensor network wavefunctions of the ferromagnetic and antiferromagnetic models, leading to exactly the same variational energy. Therefore, once one optimizes the variational parameters with the ferromagnetic model, the ansatz for the antiferromagnetic model is obtained without additional optimization\,(or vice versa). The magnetic field effect on the KSL is similar to that of the $S$=$1/2$ model. That is, the antiferromagnetic KSL is much more robust than the ferromagnetic KSL against the field. We believe that our construction is applicable to the general spin-$S$ Kitaev model.

We thank Ilia Khait and Hae-Young Kee for helpful discussions. The computation in the present work was executed on computers at the Supercomputer Center, ISSP, University of Tokyo, and also on K-computer and Information Initiative Center, Hokkaido University(project-ID: hp190196). N.K.'s work is funded by MEXT KAKENHI No.19H01809. H.-Y.L. was supported by MEXT as ``Exploratory Challenge on Post-K computer"\,(Frontiers of Basic Science: Challenging the Limits). Y.B.K. was supported by the NSERC of Canada and the Killam Research Fellowship of the Canada Council for the Arts.

\bibliographystyle{apsrev}
\bibliography{reference.bib}

\begin{thebibliography}{40}
\expandafter\ifx\csname natexlab\endcsname\relax\def\natexlab#1{#1}\fi
\expandafter\ifx\csname bibnamefont\endcsname\relax
  \def\bibnamefont#1{#1}\fi
\expandafter\ifx\csname bibfnamefont\endcsname\relax
  \def\bibfnamefont#1{#1}\fi
\expandafter\ifx\csname citenamefont\endcsname\relax
  \def\citenamefont#1{#1}\fi
\expandafter\ifx\csname url\endcsname\relax
  \def\url#1{\texttt{#1}}\fi
\expandafter\ifx\csname urlprefix\endcsname\relax\def\urlprefix{URL }\fi
\providecommand{\bibinfo}[2]{#2}
\providecommand{\eprint}[2][]{\url{#2}}

\bibitem[{\citenamefont{Kitaev}(2006)}]{Kitaev2006}
\bibinfo{author}{\bibfnamefont{A.}~\bibnamefont{Kitaev}},
  \bibinfo{journal}{Annals of Physics} \textbf{\bibinfo{volume}{321}},
  \bibinfo{pages}{2} (\bibinfo{year}{2006}), ISSN \bibinfo{issn}{00034916},
  \eprint{0506438}.

\bibitem[{\citenamefont{Khaliullin}(2005)}]{Khaliullin2005}
\bibinfo{author}{\bibfnamefont{G.}~\bibnamefont{Khaliullin}},
  \bibinfo{journal}{Progress of Theoretical Physics Supplement}
  \textbf{\bibinfo{volume}{160}}, \bibinfo{pages}{155} (\bibinfo{year}{2005}).

\bibitem[{\citenamefont{Jackeli and Khaliullin}(2009)}]{Jackeli2009}
\bibinfo{author}{\bibfnamefont{G.}~\bibnamefont{Jackeli}} \bibnamefont{and}
  \bibinfo{author}{\bibfnamefont{G.}~\bibnamefont{Khaliullin}},
  \bibinfo{journal}{Phys. Rev. Lett.} \textbf{\bibinfo{volume}{102}},
  \bibinfo{pages}{017205} (\bibinfo{year}{2009}),
  \urlprefix\url{https://link.aps.org/doi/10.1103/PhysRevLett.102.017205}.

\bibitem[{\citenamefont{Plumb et~al.}(2014)\citenamefont{Plumb, Clancy,
  Sandilands, Shankar, Hu, Burch, Kee, and Kim}}]{Plumb2014}
\bibinfo{author}{\bibfnamefont{K.~W.} \bibnamefont{Plumb}},
  \bibinfo{author}{\bibfnamefont{J.~P.} \bibnamefont{Clancy}},
  \bibinfo{author}{\bibfnamefont{L.~J.} \bibnamefont{Sandilands}},
  \bibinfo{author}{\bibfnamefont{V.~V.} \bibnamefont{Shankar}},
  \bibinfo{author}{\bibfnamefont{Y.~F.} \bibnamefont{Hu}},
  \bibinfo{author}{\bibfnamefont{K.~S.} \bibnamefont{Burch}},
  \bibinfo{author}{\bibfnamefont{H.-Y.} \bibnamefont{Kee}}, \bibnamefont{and}
  \bibinfo{author}{\bibfnamefont{Y.-J.} \bibnamefont{Kim}},
  \bibinfo{journal}{Phys. Rev. B} \textbf{\bibinfo{volume}{90}},
  \bibinfo{pages}{041112} (\bibinfo{year}{2014}),
  \urlprefix\url{https://link.aps.org/doi/10.1103/PhysRevB.90.041112}.

\bibitem[{\citenamefont{Sears et~al.}(2015)\citenamefont{Sears, Songvilay,
  Plumb, Clancy, Qiu, Zhao, Parshall, and Kim}}]{Sears15}
\bibinfo{author}{\bibfnamefont{J.~A.} \bibnamefont{Sears}},
  \bibinfo{author}{\bibfnamefont{M.}~\bibnamefont{Songvilay}},
  \bibinfo{author}{\bibfnamefont{K.~W.} \bibnamefont{Plumb}},
  \bibinfo{author}{\bibfnamefont{J.~P.} \bibnamefont{Clancy}},
  \bibinfo{author}{\bibfnamefont{Y.}~\bibnamefont{Qiu}},
  \bibinfo{author}{\bibfnamefont{Y.}~\bibnamefont{Zhao}},
  \bibinfo{author}{\bibfnamefont{D.}~\bibnamefont{Parshall}}, \bibnamefont{and}
  \bibinfo{author}{\bibfnamefont{Y.-J.} \bibnamefont{Kim}},
  \bibinfo{journal}{Phys. Rev. B} \textbf{\bibinfo{volume}{91}},
  \bibinfo{pages}{144420} (\bibinfo{year}{2015}),
  \urlprefix\url{https://link.aps.org/doi/10.1103/PhysRevB.91.144420}.

\bibitem[{\citenamefont{Johnson et~al.}(2015)\citenamefont{Johnson, Williams,
  Haghighirad, Singleton, Zapf, Manuel, Mazin, Li, Jeschke, Valent\'{\i}
  et~al.}}]{Johnson15}
\bibinfo{author}{\bibfnamefont{R.~D.} \bibnamefont{Johnson}},
  \bibinfo{author}{\bibfnamefont{S.~C.} \bibnamefont{Williams}},
  \bibinfo{author}{\bibfnamefont{A.~A.} \bibnamefont{Haghighirad}},
  \bibinfo{author}{\bibfnamefont{J.}~\bibnamefont{Singleton}},
  \bibinfo{author}{\bibfnamefont{V.}~\bibnamefont{Zapf}},
  \bibinfo{author}{\bibfnamefont{P.}~\bibnamefont{Manuel}},
  \bibinfo{author}{\bibfnamefont{I.~I.} \bibnamefont{Mazin}},
  \bibinfo{author}{\bibfnamefont{Y.}~\bibnamefont{Li}},
  \bibinfo{author}{\bibfnamefont{H.~O.} \bibnamefont{Jeschke}},
  \bibinfo{author}{\bibfnamefont{R.}~\bibnamefont{Valent\'{\i}}},
  \bibnamefont{et~al.}, \bibinfo{journal}{Phys. Rev. B}
  \textbf{\bibinfo{volume}{92}}, \bibinfo{pages}{235119}
  (\bibinfo{year}{2015}),
  \urlprefix\url{https://link.aps.org/doi/10.1103/PhysRevB.92.235119}.

\bibitem[{\citenamefont{Kim et~al.}(2015)\citenamefont{Kim, V., Catuneanu, and
  Kee}}]{Kim15}
\bibinfo{author}{\bibfnamefont{H.-S.} \bibnamefont{Kim}},
  \bibinfo{author}{\bibfnamefont{V.~S.} \bibnamefont{V.}},
  \bibinfo{author}{\bibfnamefont{A.}~\bibnamefont{Catuneanu}},
  \bibnamefont{and} \bibinfo{author}{\bibfnamefont{H.-Y.} \bibnamefont{Kee}},
  \bibinfo{journal}{Phys. Rev. B} \textbf{\bibinfo{volume}{91}},
  \bibinfo{pages}{241110} (\bibinfo{year}{2015}),
  \urlprefix\url{https://link.aps.org/doi/10.1103/PhysRevB.91.241110}.

\bibitem[{\citenamefont{Kim and Kee}(2016)}]{Kim16}
\bibinfo{author}{\bibfnamefont{H.-S.} \bibnamefont{Kim}} \bibnamefont{and}
  \bibinfo{author}{\bibfnamefont{H.-Y.} \bibnamefont{Kee}},
  \bibinfo{journal}{Phys. Rev. B} \textbf{\bibinfo{volume}{93}},
  \bibinfo{pages}{155143} (\bibinfo{year}{2016}),
  \urlprefix\url{https://link.aps.org/doi/10.1103/PhysRevB.93.155143}.

\bibitem[{\citenamefont{Yadav et~al.}(2016)\citenamefont{Yadav, Bogdanov,
  Katukuri, Nishimoto, Van Den~Brink, and Hozoi}}]{Yadav16}
\bibinfo{author}{\bibfnamefont{R.}~\bibnamefont{Yadav}},
  \bibinfo{author}{\bibfnamefont{N.~A.} \bibnamefont{Bogdanov}},
  \bibinfo{author}{\bibfnamefont{V.~M.} \bibnamefont{Katukuri}},
  \bibinfo{author}{\bibfnamefont{S.}~\bibnamefont{Nishimoto}},
  \bibinfo{author}{\bibfnamefont{J.}~\bibnamefont{Van Den~Brink}},
  \bibnamefont{and} \bibinfo{author}{\bibfnamefont{L.}~\bibnamefont{Hozoi}},
  \bibinfo{journal}{Scientific reports} \textbf{\bibinfo{volume}{6}},
  \bibinfo{pages}{37925} (\bibinfo{year}{2016}).

\bibitem[{\citenamefont{Zhou et~al.}(2016)\citenamefont{Zhou, Li, Waugh,
  Parham, Kim, Sears, Gomes, Kee, Kim, and Dessau}}]{Zhou2016b}
\bibinfo{author}{\bibfnamefont{X.}~\bibnamefont{Zhou}},
  \bibinfo{author}{\bibfnamefont{H.}~\bibnamefont{Li}},
  \bibinfo{author}{\bibfnamefont{J.}~\bibnamefont{Waugh}},
  \bibinfo{author}{\bibfnamefont{S.}~\bibnamefont{Parham}},
  \bibinfo{author}{\bibfnamefont{H.-S.} \bibnamefont{Kim}},
  \bibinfo{author}{\bibfnamefont{J.}~\bibnamefont{Sears}},
  \bibinfo{author}{\bibfnamefont{A.}~\bibnamefont{Gomes}},
  \bibinfo{author}{\bibfnamefont{H.-Y.} \bibnamefont{Kee}},
  \bibinfo{author}{\bibfnamefont{Y.-J.} \bibnamefont{Kim}}, \bibnamefont{and}
  \bibinfo{author}{\bibfnamefont{D.}~\bibnamefont{Dessau}},
  \bibinfo{journal}{Physical Review B} \textbf{\bibinfo{volume}{94}},
  \bibinfo{pages}{161106} (\bibinfo{year}{2016}).

\bibitem[{\citenamefont{Banerjee et~al.}(2016)\citenamefont{Banerjee, Bridges,
  Yan, Aczel, Li, Stone, Granroth, Lumsden, Yiu, Knolle et~al.}}]{Banerjee2016}
\bibinfo{author}{\bibfnamefont{A.}~\bibnamefont{Banerjee}},
  \bibinfo{author}{\bibfnamefont{C.}~\bibnamefont{Bridges}},
  \bibinfo{author}{\bibfnamefont{J.-Q.} \bibnamefont{Yan}},
  \bibinfo{author}{\bibfnamefont{A.}~\bibnamefont{Aczel}},
  \bibinfo{author}{\bibfnamefont{L.}~\bibnamefont{Li}},
  \bibinfo{author}{\bibfnamefont{M.}~\bibnamefont{Stone}},
  \bibinfo{author}{\bibfnamefont{G.}~\bibnamefont{Granroth}},
  \bibinfo{author}{\bibfnamefont{M.}~\bibnamefont{Lumsden}},
  \bibinfo{author}{\bibfnamefont{Y.}~\bibnamefont{Yiu}},
  \bibinfo{author}{\bibfnamefont{J.}~\bibnamefont{Knolle}},
  \bibnamefont{et~al.}, \bibinfo{journal}{Nature materials}
  \textbf{\bibinfo{volume}{15}}, \bibinfo{pages}{733} (\bibinfo{year}{2016}).

\bibitem[{\citenamefont{Sandilands et~al.}(2016)\citenamefont{Sandilands, Tian,
  Reijnders, Kim, Plumb, Kim, Kee, and Burch}}]{Luke16}
\bibinfo{author}{\bibfnamefont{L.~J.} \bibnamefont{Sandilands}},
  \bibinfo{author}{\bibfnamefont{Y.}~\bibnamefont{Tian}},
  \bibinfo{author}{\bibfnamefont{A.~A.} \bibnamefont{Reijnders}},
  \bibinfo{author}{\bibfnamefont{H.-S.} \bibnamefont{Kim}},
  \bibinfo{author}{\bibfnamefont{K.~W.} \bibnamefont{Plumb}},
  \bibinfo{author}{\bibfnamefont{Y.-J.} \bibnamefont{Kim}},
  \bibinfo{author}{\bibfnamefont{H.-Y.} \bibnamefont{Kee}}, \bibnamefont{and}
  \bibinfo{author}{\bibfnamefont{K.~S.} \bibnamefont{Burch}},
  \bibinfo{journal}{Phys. Rev. B} \textbf{\bibinfo{volume}{93}},
  \bibinfo{pages}{075144} (\bibinfo{year}{2016}),
  \urlprefix\url{https://link.aps.org/doi/10.1103/PhysRevB.93.075144}.

\bibitem[{\citenamefont{Sinn et~al.}(2016)\citenamefont{Sinn, Kim, Kim, Lee,
  Won, Oh, Han, Chang, Hur, Sato et~al.}}]{Sinn16}
\bibinfo{author}{\bibfnamefont{S.}~\bibnamefont{Sinn}},
  \bibinfo{author}{\bibfnamefont{C.~H.} \bibnamefont{Kim}},
  \bibinfo{author}{\bibfnamefont{B.~H.} \bibnamefont{Kim}},
  \bibinfo{author}{\bibfnamefont{K.~D.} \bibnamefont{Lee}},
  \bibinfo{author}{\bibfnamefont{C.~J.} \bibnamefont{Won}},
  \bibinfo{author}{\bibfnamefont{J.~S.} \bibnamefont{Oh}},
  \bibinfo{author}{\bibfnamefont{M.}~\bibnamefont{Han}},
  \bibinfo{author}{\bibfnamefont{Y.~J.} \bibnamefont{Chang}},
  \bibinfo{author}{\bibfnamefont{N.}~\bibnamefont{Hur}},
  \bibinfo{author}{\bibfnamefont{H.}~\bibnamefont{Sato}}, \bibnamefont{et~al.},
  \bibinfo{journal}{Scientific reports} \textbf{\bibinfo{volume}{6}},
  \bibinfo{pages}{39544} (\bibinfo{year}{2016}).

\bibitem[{\citenamefont{Winter et~al.}(2016)\citenamefont{Winter, Li, Jeschke,
  and Valent\'{\i}}}]{Winter16}
\bibinfo{author}{\bibfnamefont{S.~M.} \bibnamefont{Winter}},
  \bibinfo{author}{\bibfnamefont{Y.}~\bibnamefont{Li}},
  \bibinfo{author}{\bibfnamefont{H.~O.} \bibnamefont{Jeschke}},
  \bibnamefont{and}
  \bibinfo{author}{\bibfnamefont{R.}~\bibnamefont{Valent\'{\i}}},
  \bibinfo{journal}{Phys. Rev. B} \textbf{\bibinfo{volume}{93}},
  \bibinfo{pages}{214431} (\bibinfo{year}{2016}),
  \urlprefix\url{https://link.aps.org/doi/10.1103/PhysRevB.93.214431}.

\bibitem[{\citenamefont{Leahy et~al.}(2017)\citenamefont{Leahy, Pocs,
  Siegfried, Graf, Do, Choi, Normand, and Lee}}]{Leahy17a}
\bibinfo{author}{\bibfnamefont{I.~A.} \bibnamefont{Leahy}},
  \bibinfo{author}{\bibfnamefont{C.~A.} \bibnamefont{Pocs}},
  \bibinfo{author}{\bibfnamefont{P.~E.} \bibnamefont{Siegfried}},
  \bibinfo{author}{\bibfnamefont{D.}~\bibnamefont{Graf}},
  \bibinfo{author}{\bibfnamefont{S.-H.} \bibnamefont{Do}},
  \bibinfo{author}{\bibfnamefont{K.-Y.} \bibnamefont{Choi}},
  \bibinfo{author}{\bibfnamefont{B.}~\bibnamefont{Normand}}, \bibnamefont{and}
  \bibinfo{author}{\bibfnamefont{M.}~\bibnamefont{Lee}},
  \bibinfo{journal}{Phys. Rev. Lett.} \textbf{\bibinfo{volume}{118}},
  \bibinfo{pages}{187203} (\bibinfo{year}{2017}),
  \urlprefix\url{https://link.aps.org/doi/10.1103/PhysRevLett.118.187203}.

\bibitem[{\citenamefont{Trebst}(2017)}]{Trebst2017}
\bibinfo{author}{\bibfnamefont{S.}~\bibnamefont{Trebst}},
  \bibinfo{journal}{arXiv preprint arXiv:1701.07056}  (\bibinfo{year}{2017}).

\bibitem[{\citenamefont{Banerjee et~al.}(2017)\citenamefont{Banerjee, Yan,
  Knolle, Bridges, Stone, Lumsden, Mandrus, Tennant, Moessner, and
  Nagler}}]{Banerjee17}
\bibinfo{author}{\bibfnamefont{A.}~\bibnamefont{Banerjee}},
  \bibinfo{author}{\bibfnamefont{J.}~\bibnamefont{Yan}},
  \bibinfo{author}{\bibfnamefont{J.}~\bibnamefont{Knolle}},
  \bibinfo{author}{\bibfnamefont{C.~A.} \bibnamefont{Bridges}},
  \bibinfo{author}{\bibfnamefont{M.~B.} \bibnamefont{Stone}},
  \bibinfo{author}{\bibfnamefont{M.~D.} \bibnamefont{Lumsden}},
  \bibinfo{author}{\bibfnamefont{D.~G.} \bibnamefont{Mandrus}},
  \bibinfo{author}{\bibfnamefont{D.~A.} \bibnamefont{Tennant}},
  \bibinfo{author}{\bibfnamefont{R.}~\bibnamefont{Moessner}}, \bibnamefont{and}
  \bibinfo{author}{\bibfnamefont{S.~E.} \bibnamefont{Nagler}},
  \bibinfo{journal}{Science} \textbf{\bibinfo{volume}{356}},
  \bibinfo{pages}{1055} (\bibinfo{year}{2017}), ISSN \bibinfo{issn}{0036-8075},
  \eprint{https://science.sciencemag.org/content/356/6342/1055.full.pdf},
  \urlprefix\url{https://science.sciencemag.org/content/356/6342/1055}.

\bibitem[{\citenamefont{Catuneanu et~al.}(2018)\citenamefont{Catuneanu, Yamaji,
  Wachtel, Kim, and Kee}}]{Catuneanu18}
\bibinfo{author}{\bibfnamefont{A.}~\bibnamefont{Catuneanu}},
  \bibinfo{author}{\bibfnamefont{Y.}~\bibnamefont{Yamaji}},
  \bibinfo{author}{\bibfnamefont{G.}~\bibnamefont{Wachtel}},
  \bibinfo{author}{\bibfnamefont{Y.~B.} \bibnamefont{Kim}}, \bibnamefont{and}
  \bibinfo{author}{\bibfnamefont{H.-Y.} \bibnamefont{Kee}},
  \bibinfo{journal}{npj Quantum Materials} \textbf{\bibinfo{volume}{3}},
  \bibinfo{pages}{23} (\bibinfo{year}{2018}).

\bibitem[{\citenamefont{Gohlke et~al.}(2018)\citenamefont{Gohlke, Wachtel,
  Yamaji, Pollmann, and Kim}}]{Gohlke18}
\bibinfo{author}{\bibfnamefont{M.}~\bibnamefont{Gohlke}},
  \bibinfo{author}{\bibfnamefont{G.}~\bibnamefont{Wachtel}},
  \bibinfo{author}{\bibfnamefont{Y.}~\bibnamefont{Yamaji}},
  \bibinfo{author}{\bibfnamefont{F.}~\bibnamefont{Pollmann}}, \bibnamefont{and}
  \bibinfo{author}{\bibfnamefont{Y.~B.} \bibnamefont{Kim}},
  \bibinfo{journal}{Phys. Rev. B} \textbf{\bibinfo{volume}{97}},
  \bibinfo{pages}{075126} (\bibinfo{year}{2018}),
  \urlprefix\url{https://link.aps.org/doi/10.1103/PhysRevB.97.075126}.

\bibitem[{\citenamefont{Winter et~al.}(2018)\citenamefont{Winter, Riedl, Kaib,
  Coldea, and Valent\'{\i}}}]{Winter18}
\bibinfo{author}{\bibfnamefont{S.~M.} \bibnamefont{Winter}},
  \bibinfo{author}{\bibfnamefont{K.}~\bibnamefont{Riedl}},
  \bibinfo{author}{\bibfnamefont{D.}~\bibnamefont{Kaib}},
  \bibinfo{author}{\bibfnamefont{R.}~\bibnamefont{Coldea}}, \bibnamefont{and}
  \bibinfo{author}{\bibfnamefont{R.}~\bibnamefont{Valent\'{\i}}},
  \bibinfo{journal}{Phys. Rev. Lett.} \textbf{\bibinfo{volume}{120}},
  \bibinfo{pages}{077203} (\bibinfo{year}{2018}),
  \urlprefix\url{https://link.aps.org/doi/10.1103/PhysRevLett.120.077203}.

\bibitem[{\citenamefont{Banerjee et~al.}(2018)\citenamefont{Banerjee,
  Lampen-Kelley, Knolle, Balz, Aczel, Winn, Liu, Pajerowski, Yan, Bridges
  et~al.}}]{Banerjee2018}
\bibinfo{author}{\bibfnamefont{A.}~\bibnamefont{Banerjee}},
  \bibinfo{author}{\bibfnamefont{P.}~\bibnamefont{Lampen-Kelley}},
  \bibinfo{author}{\bibfnamefont{J.}~\bibnamefont{Knolle}},
  \bibinfo{author}{\bibfnamefont{C.}~\bibnamefont{Balz}},
  \bibinfo{author}{\bibfnamefont{A.~A.} \bibnamefont{Aczel}},
  \bibinfo{author}{\bibfnamefont{B.}~\bibnamefont{Winn}},
  \bibinfo{author}{\bibfnamefont{Y.}~\bibnamefont{Liu}},
  \bibinfo{author}{\bibfnamefont{D.}~\bibnamefont{Pajerowski}},
  \bibinfo{author}{\bibfnamefont{J.}~\bibnamefont{Yan}},
  \bibinfo{author}{\bibfnamefont{C.~A.} \bibnamefont{Bridges}},
  \bibnamefont{et~al.}, \bibinfo{journal}{npj Quantum Materials}
  \textbf{\bibinfo{volume}{3}}, \bibinfo{pages}{8} (\bibinfo{year}{2018}).

\bibitem[{\citenamefont{Balz et~al.}(2019)\citenamefont{Balz, Lampen-Kelley,
  Banerjee, Yan, Lu, Hu, Yadav, Takano, Liu, Tennant et~al.}}]{Balz19}
\bibinfo{author}{\bibfnamefont{C.}~\bibnamefont{Balz}},
  \bibinfo{author}{\bibfnamefont{P.}~\bibnamefont{Lampen-Kelley}},
  \bibinfo{author}{\bibfnamefont{A.}~\bibnamefont{Banerjee}},
  \bibinfo{author}{\bibfnamefont{J.}~\bibnamefont{Yan}},
  \bibinfo{author}{\bibfnamefont{Z.}~\bibnamefont{Lu}},
  \bibinfo{author}{\bibfnamefont{X.}~\bibnamefont{Hu}},
  \bibinfo{author}{\bibfnamefont{S.~M.} \bibnamefont{Yadav}},
  \bibinfo{author}{\bibfnamefont{Y.}~\bibnamefont{Takano}},
  \bibinfo{author}{\bibfnamefont{Y.}~\bibnamefont{Liu}},
  \bibinfo{author}{\bibfnamefont{D.~A.} \bibnamefont{Tennant}},
  \bibnamefont{et~al.}, \bibinfo{journal}{Phys. Rev. B}
  \textbf{\bibinfo{volume}{100}}, \bibinfo{pages}{060405}
  (\bibinfo{year}{2019}),
  \urlprefix\url{https://link.aps.org/doi/10.1103/PhysRevB.100.060405}.

\bibitem[{\citenamefont{Wang et~al.}(2019)\citenamefont{Wang, Normand, and
  Liu}}]{Wang19}
\bibinfo{author}{\bibfnamefont{J.}~\bibnamefont{Wang}},
  \bibinfo{author}{\bibfnamefont{B.}~\bibnamefont{Normand}}, \bibnamefont{and}
  \bibinfo{author}{\bibfnamefont{Z.-X.} \bibnamefont{Liu}},
  \bibinfo{journal}{arXiv preprint arXiv:1903.10026}  (\bibinfo{year}{2019}).

\bibitem[{\citenamefont{Stavropoulos et~al.}(2019)\citenamefont{Stavropoulos,
  Pereira, and Kee}}]{Peter19}
\bibinfo{author}{\bibfnamefont{P.~P.} \bibnamefont{Stavropoulos}},
  \bibinfo{author}{\bibfnamefont{D.}~\bibnamefont{Pereira}}, \bibnamefont{and}
  \bibinfo{author}{\bibfnamefont{H.-Y.} \bibnamefont{Kee}},
  \bibinfo{journal}{Phys. Rev. Lett.} \textbf{\bibinfo{volume}{123}},
  \bibinfo{pages}{037203} (\bibinfo{year}{2019}),
  \urlprefix\url{https://link.aps.org/doi/10.1103/PhysRevLett.123.037203}.

\bibitem[{\citenamefont{Baskaran et~al.}(2008)\citenamefont{Baskaran, Sen, and
  Shankar}}]{Baskaran08}
\bibinfo{author}{\bibfnamefont{G.}~\bibnamefont{Baskaran}},
  \bibinfo{author}{\bibfnamefont{D.}~\bibnamefont{Sen}}, \bibnamefont{and}
  \bibinfo{author}{\bibfnamefont{R.}~\bibnamefont{Shankar}},
  \bibinfo{journal}{Phys. Rev. B} \textbf{\bibinfo{volume}{78}},
  \bibinfo{pages}{115116} (\bibinfo{year}{2008}),
  \urlprefix\url{https://link.aps.org/doi/10.1103/PhysRevB.78.115116}.

\bibitem[{\citenamefont{Koga et~al.}(2018)\citenamefont{Koga, Tomishige, and
  Nasu}}]{Koga18}
\bibinfo{author}{\bibfnamefont{A.}~\bibnamefont{Koga}},
  \bibinfo{author}{\bibfnamefont{H.}~\bibnamefont{Tomishige}},
  \bibnamefont{and} \bibinfo{author}{\bibfnamefont{J.}~\bibnamefont{Nasu}},
  \bibinfo{journal}{Journal of the Physical Society of Japan}
  \textbf{\bibinfo{volume}{87}}, \bibinfo{pages}{063703}
  (\bibinfo{year}{2018}), \eprint{https://doi.org/10.7566/JPSJ.87.063703},
  \urlprefix\url{https://doi.org/10.7566/JPSJ.87.063703}.

\bibitem[{\citenamefont{Oitmaa et~al.}(2018)\citenamefont{Oitmaa, Koga, and
  Singh}}]{Oitmaa18}
\bibinfo{author}{\bibfnamefont{J.}~\bibnamefont{Oitmaa}},
  \bibinfo{author}{\bibfnamefont{A.}~\bibnamefont{Koga}}, \bibnamefont{and}
  \bibinfo{author}{\bibfnamefont{R.~R.~P.} \bibnamefont{Singh}},
  \bibinfo{journal}{Phys. Rev. B} \textbf{\bibinfo{volume}{98}},
  \bibinfo{pages}{214404} (\bibinfo{year}{2018}),
  \urlprefix\url{https://link.aps.org/doi/10.1103/PhysRevB.98.214404}.

\bibitem[{\citenamefont{Lee et~al.}(2019{\natexlab{a}})\citenamefont{Lee,
  Kaneko, Okubo, and Kawashima}}]{HY19}
\bibinfo{author}{\bibfnamefont{H.-Y.} \bibnamefont{Lee}},
  \bibinfo{author}{\bibfnamefont{R.}~\bibnamefont{Kaneko}},
  \bibinfo{author}{\bibfnamefont{T.}~\bibnamefont{Okubo}}, \bibnamefont{and}
  \bibinfo{author}{\bibfnamefont{N.}~\bibnamefont{Kawashima}},
  \bibinfo{journal}{Phys. Rev. Lett.} \textbf{\bibinfo{volume}{123}},
  \bibinfo{pages}{087203} (\bibinfo{year}{2019}{\natexlab{a}}),
  \urlprefix\url{https://link.aps.org/doi/10.1103/PhysRevLett.123.087203}.

\bibitem[{\citenamefont{Lee et~al.}(2019{\natexlab{b}})\citenamefont{Lee,
  Kawashima, and Kim}}]{SM}
\bibinfo{author}{\bibfnamefont{H.-Y.} \bibnamefont{Lee}},
  \bibinfo{author}{\bibfnamefont{N.}~\bibnamefont{Kawashima}},
  \bibnamefont{and} \bibinfo{author}{\bibfnamefont{Y.~B.} \bibnamefont{Kim}},
  \emph{\bibinfo{title}{Supplemental material}}
  (\bibinfo{year}{2019}{\natexlab{b}}).

\bibitem[{\citenamefont{Nienhuis}(1982)}]{Nienhuis1982}
\bibinfo{author}{\bibfnamefont{B.}~\bibnamefont{Nienhuis}},
  \bibinfo{journal}{Physical Review Letters} \textbf{\bibinfo{volume}{49}},
  \bibinfo{pages}{1062} (\bibinfo{year}{1982}).

\bibitem[{\citenamefont{Koga and Nasu}(2019)}]{Koga19}
\bibinfo{author}{\bibfnamefont{A.}~\bibnamefont{Koga}} \bibnamefont{and}
  \bibinfo{author}{\bibfnamefont{J.}~\bibnamefont{Nasu}},
  \bibinfo{journal}{Phys. Rev. B} \textbf{\bibinfo{volume}{100}},
  \bibinfo{pages}{100404} (\bibinfo{year}{2019}),
  \urlprefix\url{https://link.aps.org/doi/10.1103/PhysRevB.100.100404}.

\bibitem[{\citenamefont{Khait et~al.}()\citenamefont{Khait, Kee, and
  Kim}}]{YB19}
\bibinfo{author}{\bibfnamefont{I.}~\bibnamefont{Khait}},
  \bibinfo{author}{\bibfnamefont{H.-Y.} \bibnamefont{Kee}}, \bibnamefont{and}
  \bibinfo{author}{\bibfnamefont{Y.~B.} \bibnamefont{Kim}}, \bibinfo{note}{in
  preparation}.

\bibitem[{\citenamefont{Jiang et~al.}(2008)\citenamefont{Jiang, Weng, and
  Xiang}}]{Tao08}
\bibinfo{author}{\bibfnamefont{H.~C.} \bibnamefont{Jiang}},
  \bibinfo{author}{\bibfnamefont{Z.~Y.} \bibnamefont{Weng}}, \bibnamefont{and}
  \bibinfo{author}{\bibfnamefont{T.}~\bibnamefont{Xiang}},
  \bibinfo{journal}{Phys. Rev. Lett.} \textbf{\bibinfo{volume}{101}},
  \bibinfo{pages}{090603} (\bibinfo{year}{2008}),
  \urlprefix\url{https://link.aps.org/doi/10.1103/PhysRevLett.101.090603}.

\bibitem[{\citenamefont{Mei et~al.}(2017)\citenamefont{Mei, Chen, He, and
  Wen}}]{Mei17}
\bibinfo{author}{\bibfnamefont{J.-W.} \bibnamefont{Mei}},
  \bibinfo{author}{\bibfnamefont{J.-Y.} \bibnamefont{Chen}},
  \bibinfo{author}{\bibfnamefont{H.}~\bibnamefont{He}}, \bibnamefont{and}
  \bibinfo{author}{\bibfnamefont{X.-G.} \bibnamefont{Wen}},
  \bibinfo{journal}{Phys. Rev. B} \textbf{\bibinfo{volume}{95}},
  \bibinfo{pages}{235107} (\bibinfo{year}{2017}),
  \urlprefix\url{https://link.aps.org/doi/10.1103/PhysRevB.95.235107}.

\bibitem[{\citenamefont{Lee et~al.}(2019{\natexlab{c}})\citenamefont{Lee,
  Kaneko, Okubo, and Kawashima}}]{HY19a}
\bibinfo{author}{\bibfnamefont{H.-Y.} \bibnamefont{Lee}},
  \bibinfo{author}{\bibfnamefont{R.}~\bibnamefont{Kaneko}},
  \bibinfo{author}{\bibfnamefont{T.}~\bibnamefont{Okubo}}, \bibnamefont{and}
  \bibinfo{author}{\bibfnamefont{N.}~\bibnamefont{Kawashima}},
  \bibinfo{journal}{arXiv preprint arXiv:1907.02268}
  (\bibinfo{year}{2019}{\natexlab{c}}).

\bibitem[{\citenamefont{Zhang et~al.}(2012)\citenamefont{Zhang, Grover, Turner,
  Oshikawa, and Vishwanath}}]{Zhang12}
\bibinfo{author}{\bibfnamefont{Y.}~\bibnamefont{Zhang}},
  \bibinfo{author}{\bibfnamefont{T.}~\bibnamefont{Grover}},
  \bibinfo{author}{\bibfnamefont{A.}~\bibnamefont{Turner}},
  \bibinfo{author}{\bibfnamefont{M.}~\bibnamefont{Oshikawa}}, \bibnamefont{and}
  \bibinfo{author}{\bibfnamefont{A.}~\bibnamefont{Vishwanath}},
  \bibinfo{journal}{Phys. Rev. B} \textbf{\bibinfo{volume}{85}},
  \bibinfo{pages}{235151} (\bibinfo{year}{2012}),
  \urlprefix\url{https://link.aps.org/doi/10.1103/PhysRevB.85.235151}.

\bibitem[{\citenamefont{Cirac et~al.}(2011)\citenamefont{Cirac, Poilblanc,
  Schuch, and Verstraete}}]{Cirac11}
\bibinfo{author}{\bibfnamefont{J.~I.} \bibnamefont{Cirac}},
  \bibinfo{author}{\bibfnamefont{D.}~\bibnamefont{Poilblanc}},
  \bibinfo{author}{\bibfnamefont{N.}~\bibnamefont{Schuch}}, \bibnamefont{and}
  \bibinfo{author}{\bibfnamefont{F.}~\bibnamefont{Verstraete}},
  \bibinfo{journal}{Phys. Rev. B} \textbf{\bibinfo{volume}{83}},
  \bibinfo{pages}{245134} (\bibinfo{year}{2011}),
  \urlprefix\url{https://link.aps.org/doi/10.1103/PhysRevB.83.245134}.

\bibitem[{\citenamefont{Zhu et~al.}(2018)\citenamefont{Zhu, Kimchi, Sheng, and
  Fu}}]{Zhu18}
\bibinfo{author}{\bibfnamefont{Z.}~\bibnamefont{Zhu}},
  \bibinfo{author}{\bibfnamefont{I.}~\bibnamefont{Kimchi}},
  \bibinfo{author}{\bibfnamefont{D.~N.} \bibnamefont{Sheng}}, \bibnamefont{and}
  \bibinfo{author}{\bibfnamefont{L.}~\bibnamefont{Fu}}, \bibinfo{journal}{Phys.
  Rev. B} \textbf{\bibinfo{volume}{97}}, \bibinfo{pages}{241110}
  (\bibinfo{year}{2018}),
  \urlprefix\url{https://link.aps.org/doi/10.1103/PhysRevB.97.241110}.

\bibitem[{\citenamefont{Hickey and Trebst}(2019)}]{Hickey19}
\bibinfo{author}{\bibfnamefont{C.}~\bibnamefont{Hickey}} \bibnamefont{and}
  \bibinfo{author}{\bibfnamefont{S.}~\bibnamefont{Trebst}},
  \bibinfo{journal}{Nature communications} \textbf{\bibinfo{volume}{10}},
  \bibinfo{pages}{530} (\bibinfo{year}{2019}).

\bibitem[{\citenamefont{Lee et~al.}(2019{\natexlab{d}})\citenamefont{Lee,
  Kaneko, Chern, Okubo, Yamaji, Kawashima, and Kim}}]{HY19b}
\bibinfo{author}{\bibfnamefont{H.-Y.} \bibnamefont{Lee}},
  \bibinfo{author}{\bibfnamefont{R.}~\bibnamefont{Kaneko}},
  \bibinfo{author}{\bibfnamefont{L.~E.} \bibnamefont{Chern}},
  \bibinfo{author}{\bibfnamefont{T.}~\bibnamefont{Okubo}},
  \bibinfo{author}{\bibfnamefont{Y.}~\bibnamefont{Yamaji}},
  \bibinfo{author}{\bibfnamefont{N.}~\bibnamefont{Kawashima}},
  \bibnamefont{and} \bibinfo{author}{\bibfnamefont{Y.~B.} \bibnamefont{Kim}},
  \bibinfo{journal}{arXiv preprint arXiv:1908.07671}
  (\bibinfo{year}{2019}{\natexlab{d}}).

\end{thebibliography}

\clearpage
\onecolumngrid
\begin{center}
\textbf{\large  Supplemental Material: Tensor network wavefunction of $S=1$ Kitaev spin liquids }
\end{center}

\setcounter{equation}{0}
\setcounter{figure}{0}
\setcounter{table}{0}
\setcounter{page}{1}

\section{ loop gas operator }
\subsection{vortex freeness}
As mentioned in the main text, the loop gas\,(LG) operator $Q_{\rm LG}$ guarantees the vortex-freeness. To show this explicitly, we use Eq.\,(2) in the main tex, i.e.,
\begin{align}
	U^x Q_{\lambda \mu \nu} = \sum_{\mu',\nu'} \sigma^x_{\mu\mu'}\sigma^x_{\nu\nu'} Q_{\lambda \mu' \nu'},\quad
	U^y Q_{\lambda \mu \nu} = \sum_{\nu',\lambda'} \sigma^x_{\nu\nu'}\sigma^x_{\lambda \lambda'} Q_{\lambda' \mu \nu'},\quad
	U^z Q_{\lambda \mu \nu} = \sum_{\lambda',\mu'} \sigma^x_{\lambda\lambda'}\sigma^x_{\mu\mu'} Q_{\lambda' \mu' \nu}, 
	\label{eq:key_relation_index}
\end{align}
which can be graphically illustrated as follow:
\begin{align}
	\includegraphics[width=0.8\textwidth]{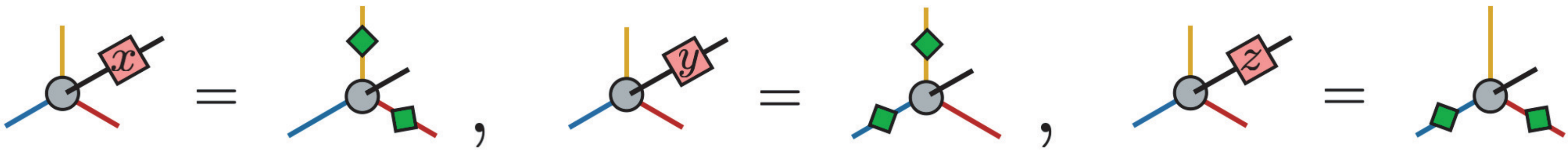},	
	\label{eq:key_relation}
\end{align}
where $U^\gamma = e^{i\pi S^\gamma}$.
In other words, applying $U^\gamma$ on the $Q$-tensor is identical to multiplying $\sigma^x$ on two virtual indices on the $\alpha$ and $\beta$ bonds where $(\alpha,\beta,\gamma)$ is a cyclic permutation of $(x,y,z)$. It is noteworthy that the $Q$-tensor in the $S=1/2$ LG operator obeys similar relations with (instead of $\sigma^x$) a non-Hermitian unitary matrix\cite{HY19}
\begin{align}
	v = \begin{pmatrix}
		0 & i \\ 1 & 0
	\end{pmatrix}	
\end{align}
and its conjugate should be applied\,(see Eq.\,(3) in Ref.\,\cite{HY19}). The vortex-freeness can be easily seen from Eq.\,\eqref{eq:key_relation}. Applying the flux operator $W_p = \exp\left[ i\pi(S_0^x + S_1^y + S_2^z + S_3^x + S_4^y + S_5^z) \right]$ on the LG operator does not affect. To be more precise, one can illustrate the application of $W_p$ on $Q_{\rm LG}$ as follows
\begin{align}
	\includegraphics[width=0.5\textwidth]{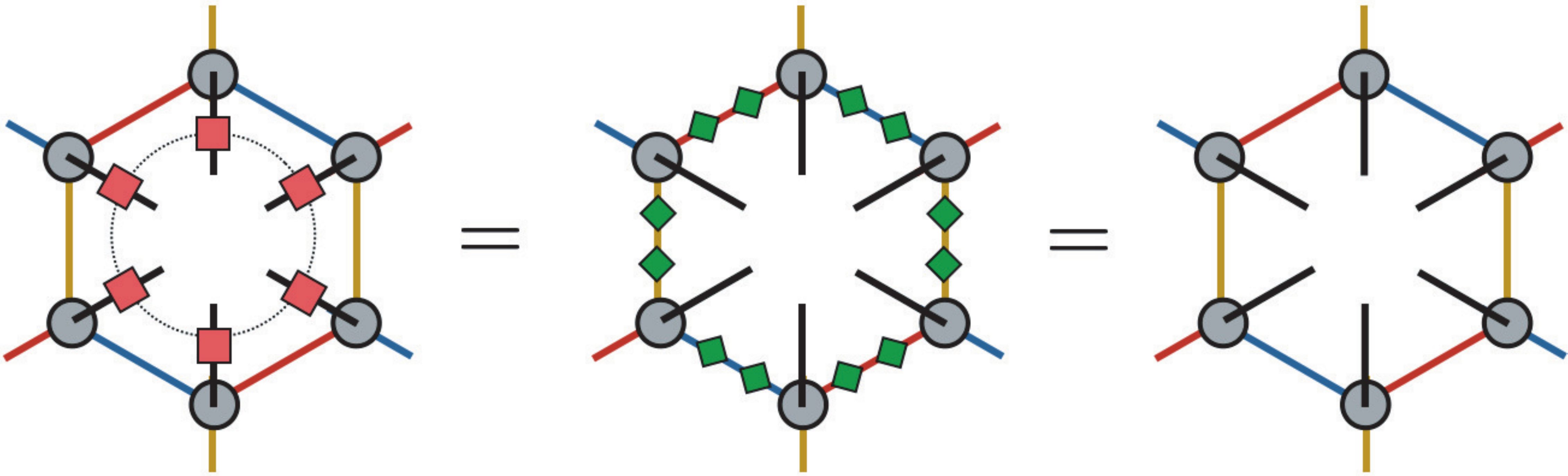}.	
	\label{eq:vortex_free}
\end{align}
where the red square denotes $U^\gamma$, and Eq.\,\eqref{eq:key_relation} and $(\sigma^x)^2 = 1$ are used for the first and second equalities. Therefore, any quantum state $|\psi\rangle = Q_{\rm LG} |0\rangle$ is vortex-free, i.e., $W_p |\psi\rangle = +|\psi\rangle$. 

\subsection{Invariant gauge group}
\label{sec:IGG}

One can also easily show that the LG operator is invariant under the $Z_2$-invariant gauge group $\{ \mathbb{I}, \sigma^z \}$, which is exactly the same as the one of the $S=1/2$ LG state. In other words, the local tensor satisfies

\begin{align}
	\includegraphics[width=0.3\textwidth]{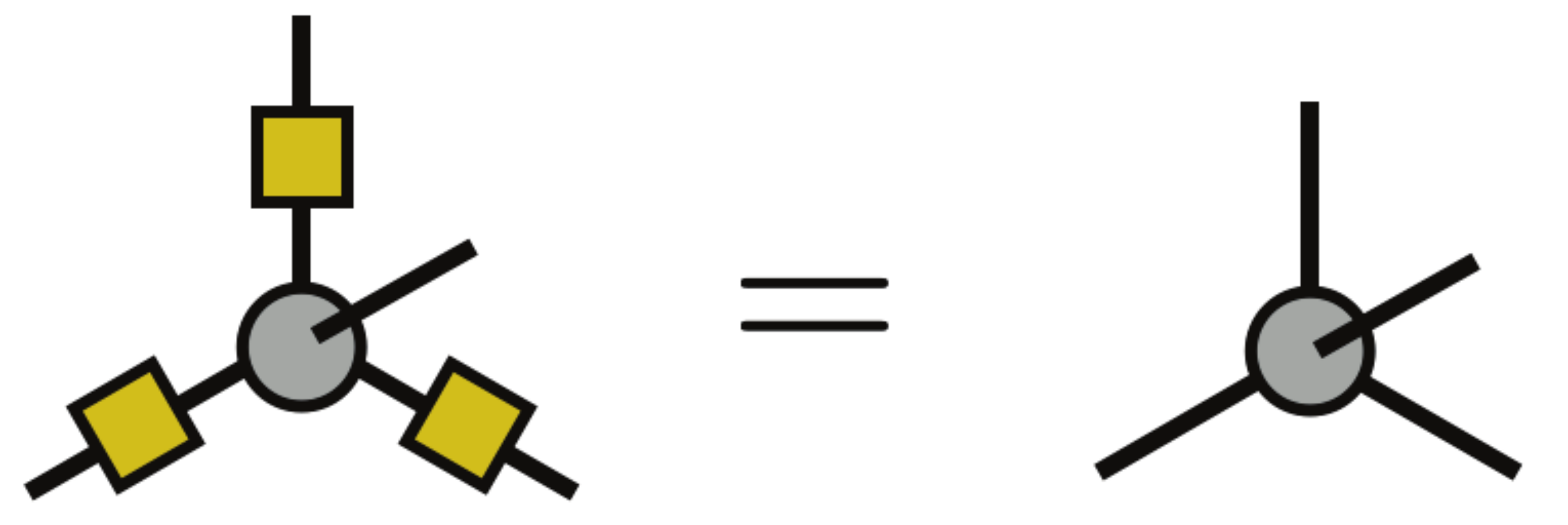},
\end{align}
where the yellow square denotes $\sigma^z$.

\subsection{Symmetries}

At the isotropic point\,($K^x=K^y=K^z$), the Hamiltonian is invariant under the $(C_6 V_{C_6})$-, $(\sigma V_\sigma)$- and time-reversal\,(TR) transformations, where $C_6$ and $\sigma$ denote the $C_6$ lattice counterclockwise rotation and inversion along a line aligned in the $z$-bond, respectively. And, the unitary operators $V_{C_6} = \exp \left( i \frac{4\pi}{3\sqrt{3}} (S^x + S^y + S^z) \right)$ and $V_\sigma = \exp \left( i \frac{\pi}{2} S^z \right)$ transform the spin operators as follows:

\begin{align}
	V_{C_6}^\dagger (S^x, S^y, S^z)	V_{C_6} = (S^z, S^x, S^y),\quad\quad
	V_\sigma^\dagger (S^x, S^y, S^z) V_\sigma  = (S^y, -S^x, S^z).
\end{align}
Due to the bond-dependent interaction in the Kitaev model, the transformation $V_{C_6}$ rotates the bond clockwise so that the combined transformation with $C_6$, or $(C_6 V_{C_6})$, leaves the Hamiltonian invariant. Similarly, it is symmetric under the $(\sigma V_\sigma)$-transformation.

One can show that the loop gas and dimer gas operators are also invariant under those transformations. Note that $V_{C_6}$ and $V_\sigma$ transform the $U^\gamma$-operator in a similar fashion:

\begin{align}
	V_{C_6}^\dagger (U^x, U^y, U^z)	V_{C_6} = (U^z, U^x, U^y),\quad\quad
	V_\sigma^\dagger (U^x, U^y, U^z) V_\sigma  = (U^y, U^x, U^z).
\end{align}
Therefore, the $Q$-tensor is transformed as follows:

\begin{align}
	V_{C_6}^\dagger Q_{\lambda \mu \nu} V_{C_6} = Q_{\mu \nu \lambda},\quad\quad
	V_\sigma^\dagger Q_{\lambda \mu \nu} V_\sigma = Q_{\mu \lambda \nu}.
\end{align}
Note that the $C_6$-rotation and $\sigma$-inversion permute the virtual indices of the tensor in the following way:

\begin{align}
	C_6 \circ T_{\lambda \mu \nu} = T_{\nu \lambda \mu },\quad\quad
	\sigma \circ T_{\lambda \mu \nu} = T_{\mu \lambda \nu}.
\end{align}
Therefore, the combined transformations $(C_6 V_{C_6})$ and $(\sigma V_\sigma)$ leave the $Q$-tensor invariant, and thus the resulting loop gas operator, too. Since the TR-transformation acts trivially on $U^\gamma$, i.e., $\mathcal{T}^{-1} U^\gamma \mathcal{T} = U^\gamma$, it leaves the loop gas operator intact. In a similar way, one can show that the dimer gas operator is symmetric under those transformations.

\section{ topological property }
\subsection{Global flux on the loop gas\,(LG) and string gas\,(SG) states}

On a compact geometry, the Hamiltonian commutes with the global flux operator $W_\mathcal{P} \equiv \prod_{i\in \mathcal{P}} U^{\gamma_i} $ where $\mathcal{P}$ is a non-contractible path winding the system. The eigenvalue $\pm 1$ of such global flux operator can be used to characterize the topological sector. On the torus geometry, we are able to define two global flux operators, say $W_h = \prod_{i\in h} U^z$ and $W_v = \prod_{i\in v} U^y$ wrapping the torus around the horizontal and vertical directions, respectively. Then, one can show that the LG state (and thus SG state) is the eigenstate of $W_h$ and $W_v$ with eigenvalue $+1$. To be more precise, let us consider the $(2\times2)$ system with the periodic boundary condition and then apply both operators on the LG state, which can be illustrated as

\begin{align}
	\includegraphics[width=0.99\textwidth]{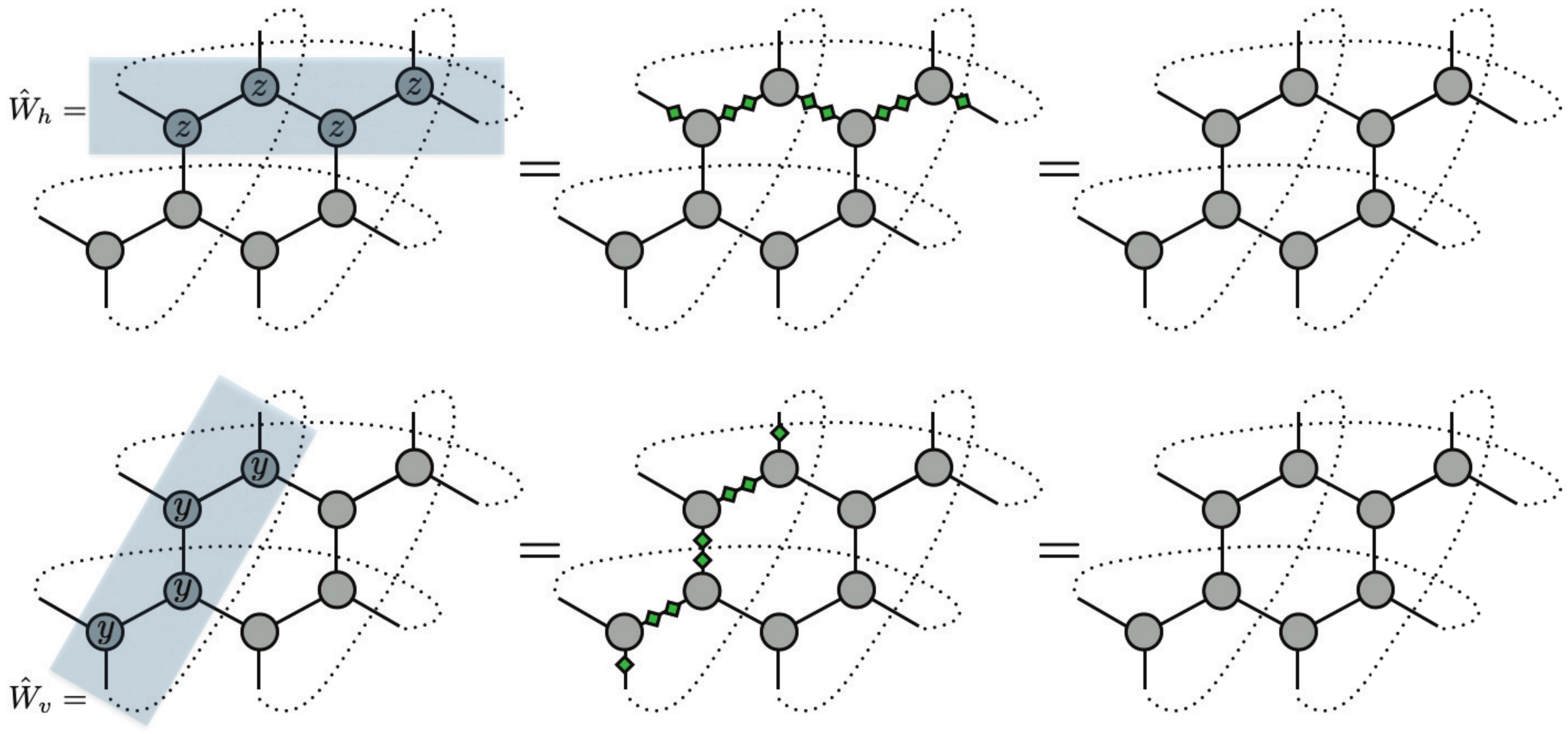},
	\label{eq:global_flux}
\end{align}
where the $z\,(y)$ on the circle stands for the multiplication of the $U^{z\,(y)}$ operator on the corresponding site. In the first equality of each equation, the equation \ref{eq:key_relation} is used, i.e. the multiplication of $U^{z\,(y)}$ is replaced by the multiplication of $\sigma^x$\,(green square) on the virtual bond. Due to $(\sigma^x)^2 = 1$, all the green squares in the second expression in each equation are cancelled, and thus $W_h |{\rm LG}\rangle = +1 |{\rm LG}\rangle  = W_v |{\rm LG}\rangle$. It indicates that our LG state on the torus geometry is already the eigenstate of both global flux operators in contrary to the $S=1/2$ LG state\cite{HY19a}. Due to this difference in the $S=1$ and $S=1/2$ LG states, the construction of the minimally entangled state\,(MES) also differs from each other.

\subsection{ Access to other global flux sector }

As shown above, the LG state obtained by the LG operator is in the $(W_h, W_v) = (+1,+1)$ sector. Note that one can access other sectors by inserting the row and column of the non-trivial element\,($g = \sigma^z$) of IGG around the system, say $\mathcal{G}_h$ and $\mathcal{G}_v$. Then, it changes the eigenvalue of the global flux operator. For instance, in the $(2\times 2)$-system, let us insert it along the vertical direction\,($\mathcal{G}_v$) and then apply the $W_h$ operator as depicted below:

\begin{align}
	\includegraphics[width=0.99\textwidth]{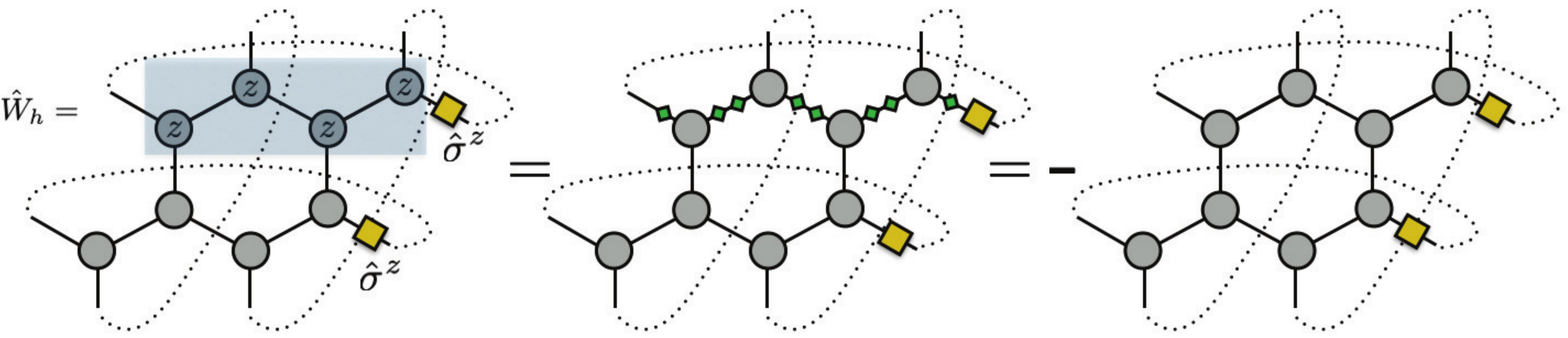}.
\end{align}
Similar to Eq.\,\eqref{eq:global_flux}, the multiplication of $U^z$ can be replaced by multiplication of $\sigma^x$\,(green square) on two virtual legs, and two successive green squares are cancelled. However, note that $\sigma^x \sigma^z \sigma^x = -\sigma^z$ appears when the $\sigma^z$ is inserted. Therefore, inserting $\mathcal{G}_v$ flips the horizontal global flux number $W_h$. In a similar fashion, one can construct the LG states living in each sector $(W_h, W_v) = (\pm, \pm)$:

\begin{align}
	\includegraphics[width=0.99\textwidth]{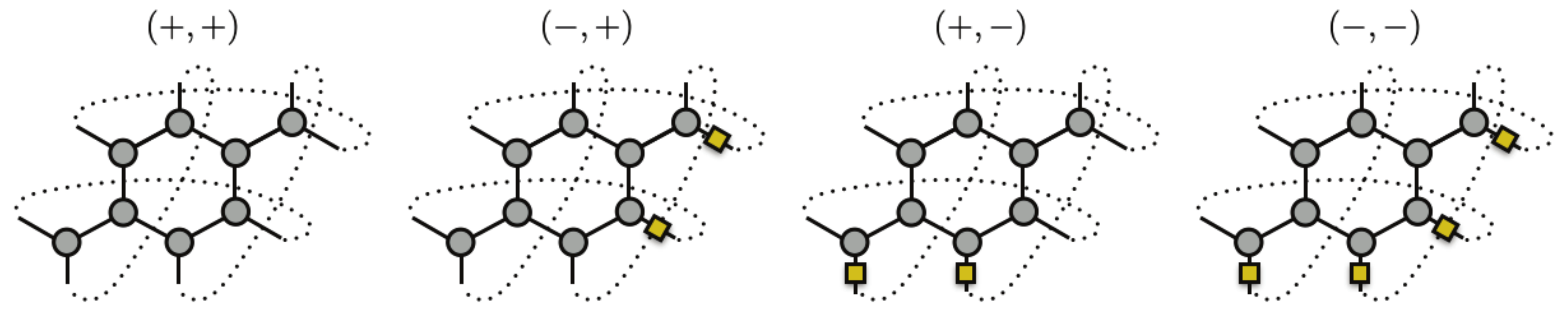}
\end{align}
which is very different from the $S=1/2$ case where, for instance, the $(+,+)$-sector is obtained by summing over all of the states above\cite{HY19a}:

\begin{align}
	\includegraphics[width=0.99\textwidth]{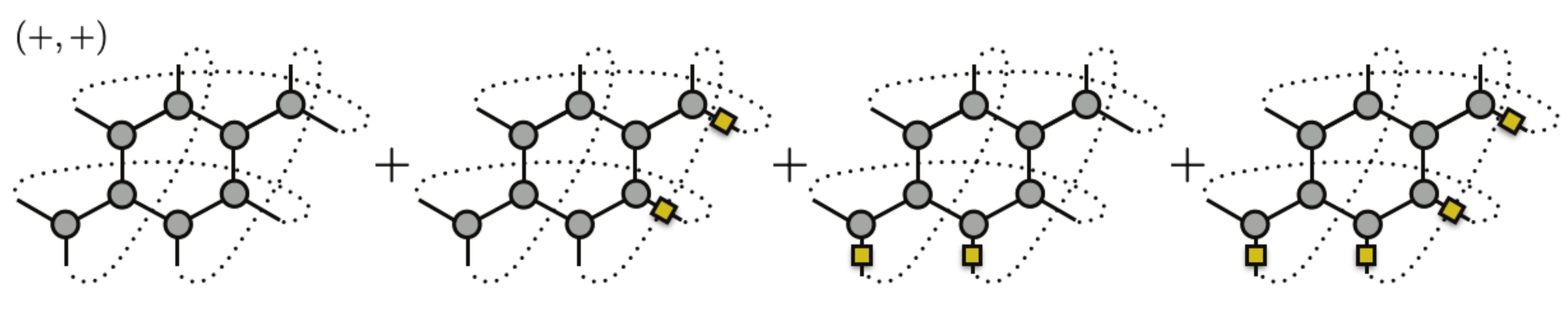}
\end{align}
In the case of the $n$-th order SG state, the non-trivial element of IGG becomes $g = \sigma^z \otimes_{i=1}^n \mathbb{I}_2 $, and the construction of each sector is exactly the same as above.

\subsection{Minimally entangled states on the infinite cylinder}

In the presence of the $Z_2$ gauge structure, one can set the MES on the infinite cylinder\,($L_x \rightarrow \infty$, $L_y$: finite) as follows\cite{Zhang12}:

\begin{align}
	|\mathbb{I}\rangle = \frac{1}{\sqrt{2}} \Big[ |(+,+)\rangle + |(+,-)\rangle \Big],\quad
	|e\rangle = \frac{1}{\sqrt{2}} \Big[ |(+,+)\rangle - |(+,-)\rangle \Big],\nonumber\\
	|m\rangle = \frac{1}{\sqrt{2}} \Big[ |(-,+)\rangle + |(-,-)\rangle \Big],\quad
	|\epsilon\rangle = \frac{1}{\sqrt{2}} \Big[ |(-,+)\rangle - |(-,-)\rangle \Big].
\end{align}
Since any flux sector can be accessed simply by inserting $\mathcal{G}_h$ and $\mathcal{G}_v$ in the original tensor network, we can construct all the minimally entangled states.

\section{ Tensor network deformation for Anti-ferromagnetic model }
\begin{figure}[!h]
	\includegraphics[width=0.9\textwidth]{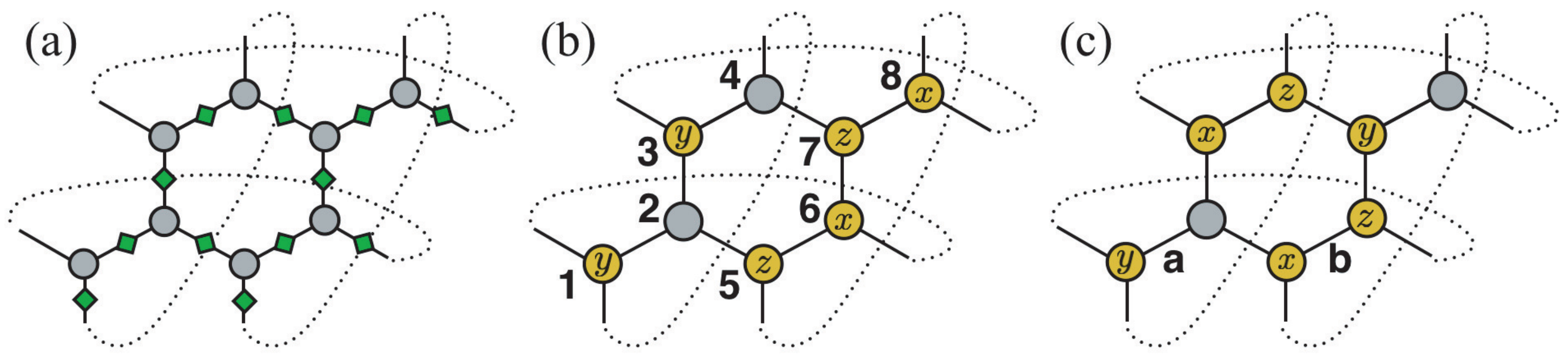}
	\caption{ Graphical representation of (a) the deformed LG operator $\widetilde{Q}_{\rm LG}$, the application of (b) $V = U^{y}_1  U^{y}_3 U^{z}_5  U^{x}_6  U^{z}_7  U^{x}_8$ and (c) $V' = U^{y}_1   U^{x}_3  U^{z}_4  U^{x}_5  U^{z}_6  U^{y}_7  $ to the original LG operator $Q_{\rm LG}$. Here, the yellow circle with $\gamma$ denotes $U^\gamma$.  }
	\label{fig:afm_deformation}
\end{figure}

As pointed out in the main text, the deformed LG state can be used for the antiferromagnetic model. First, using Eqs.\,\eqref{eq:key_relation_index} and \eqref{eq:key_relation}, one can show that the deformed LG operator depicted in Fig.\,\ref{fig:afm_deformation}\,(a) guarantees the vortex-freeness as follows. In an analogous manner to Eq.\,\eqref{eq:vortex_free}, applying the flux operator to the deformed LG operator, one obtains a similar equation

\begin{align}
	\includegraphics[width=0.6\textwidth]{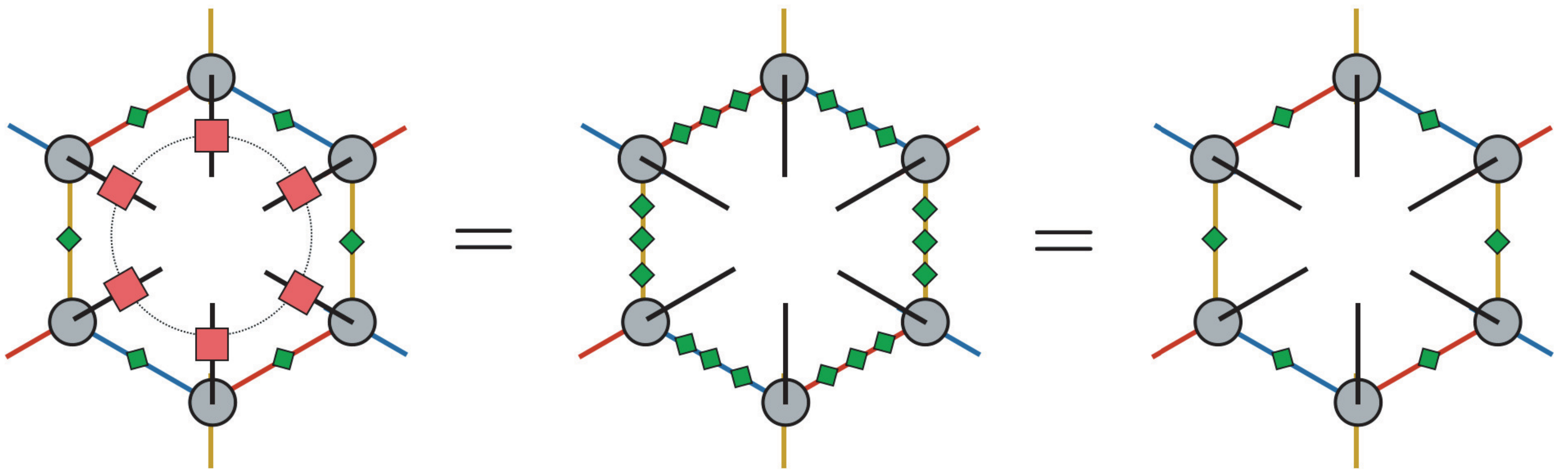},
\end{align}
where the green square denotes the $\sigma^x$-matrix, and Eq.\,\eqref{eq:key_relation} and $(\sigma^x)^3=\sigma^x$ are used in the first and second equalities, respectively. In other words, this equation implies $W_p \widetilde{Q}_{\rm LG} = \widetilde{Q}_{\rm LG}$. 

Furthermore,  using Eq.\,\eqref{eq:key_relation}, one can also easily prove that the $\widetilde{Q}_{\rm LG}$-operator can be obtained by applying some unitary operator $V$, which consists of $U^\gamma$, to the original LG operator $Q_{\rm LG}$: $\widetilde{Q}_{\rm LG} = V Q_{\rm LG}$. For example, let us apply $V = U^{y}_1  U^{y}_3 U^{z}_5  U^{x}_6  U^{z}_7  U^{x}_8$ to the original LG operator $Q_{\rm LG}$ in a $(2\times2)$-unitcell system as depicted in Fig.\,\ref{fig:afm_deformation}\,(b). Then, using Eq.\,\eqref{eq:key_relation}, one can show Fig.\,\ref{fig:afm_deformation}\,(b) is identical to Fig.\,\ref{fig:afm_deformation}\,(a), i.e., $V Q_{\rm LG} = \widetilde{Q}_{\rm LG}$. In addition, such a unitary operator $V$ is not unique, e.g., $V' Q_{\rm LG} = \widetilde{Q}_{\rm LG}$ where $V'$ is defined in Fig.\,\ref{fig:afm_deformation}\,(c) is another possibility. Note that all $V$ unitary transformation are related by the product of flux operators: $V' = (\prod_{p\in {\rm arbitrary}} W_p )V$. Finally, note that the $V$-transformation flips the sign of the Hamiltonian, i.e., $V^\dagger H V = -H$. For instance, on the x-bond $a$ in Fig.\,\ref{fig:afm_deformation}\,(c), one can easily see $V^\dagger H^x_a V = -H^x_a$ using the identities $\{ S^\gamma, U^{\gamma'} \} = 0$ for $\gamma \neq \gamma$ and $[ S^\gamma, U^{\gamma} ] = 0$. Similarly, all local Hamiltonian satisfies $V^\dagger H^\gamma_{ij} V = -H^\gamma_{ij}$ such that the full Hamiltonian obeys $V^\dagger H V = -H$. 

\begin{figure}[!h]
	\includegraphics[width=0.7\textwidth]{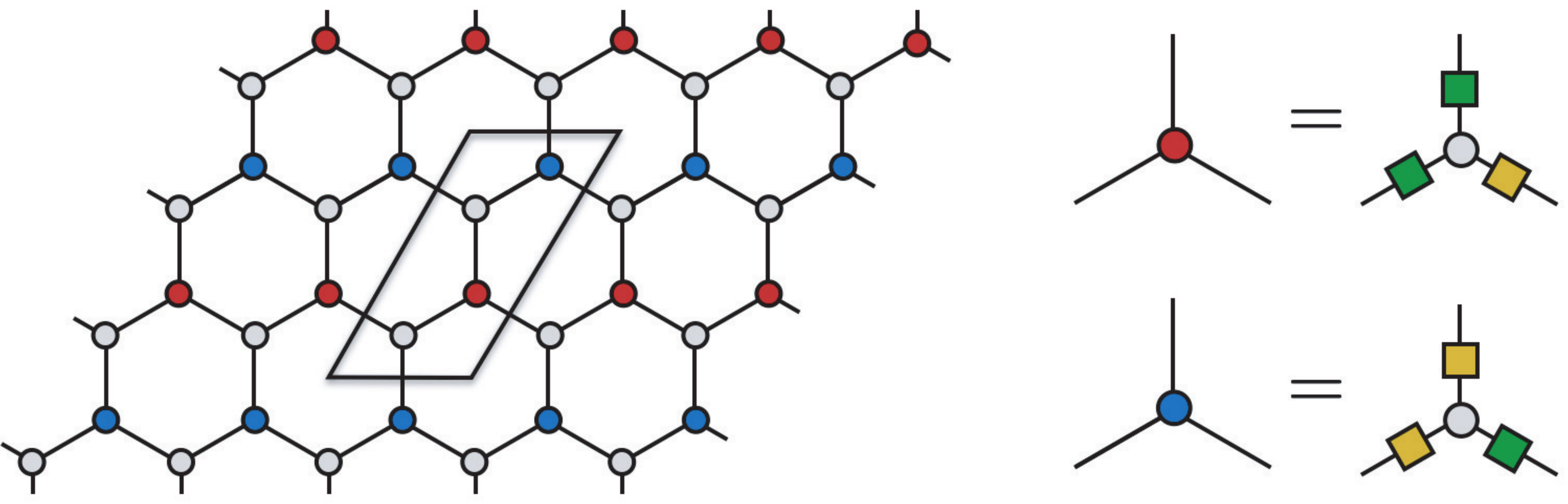}
	\caption{ Deformation of the $S$=$1/2$ LG operator for the antiferromagnetic model. Here, the grey circle denotes the original $Q$-tensor, and red and blue circles stand for the decorated tensors by matrices $v$\,(yellow square) and $v^*$\,(green square). The definition of $v$ is presented in Ref.\,\cite{HY19}.}
	\label{fig:spin_half_afm}
\end{figure}

In the case of the $S$=$1/2$, one can easily deform the LG and SG ansatz for the antiferromagnetic model. However, it requires the larger unit-cell TN, i.e., four-site unit-cell, as depicted in Fig.\,\ref{fig:spin_half_afm}.

\section{ Additional numerical results }
\subsection{ Energy landscape of the second order string gas ansatz $|\psi_{{\rm SG}_2}(\phi_1,\phi_2)\rangle$ }

\begin{figure}[!h]
	\includegraphics[width=0.4\textwidth]{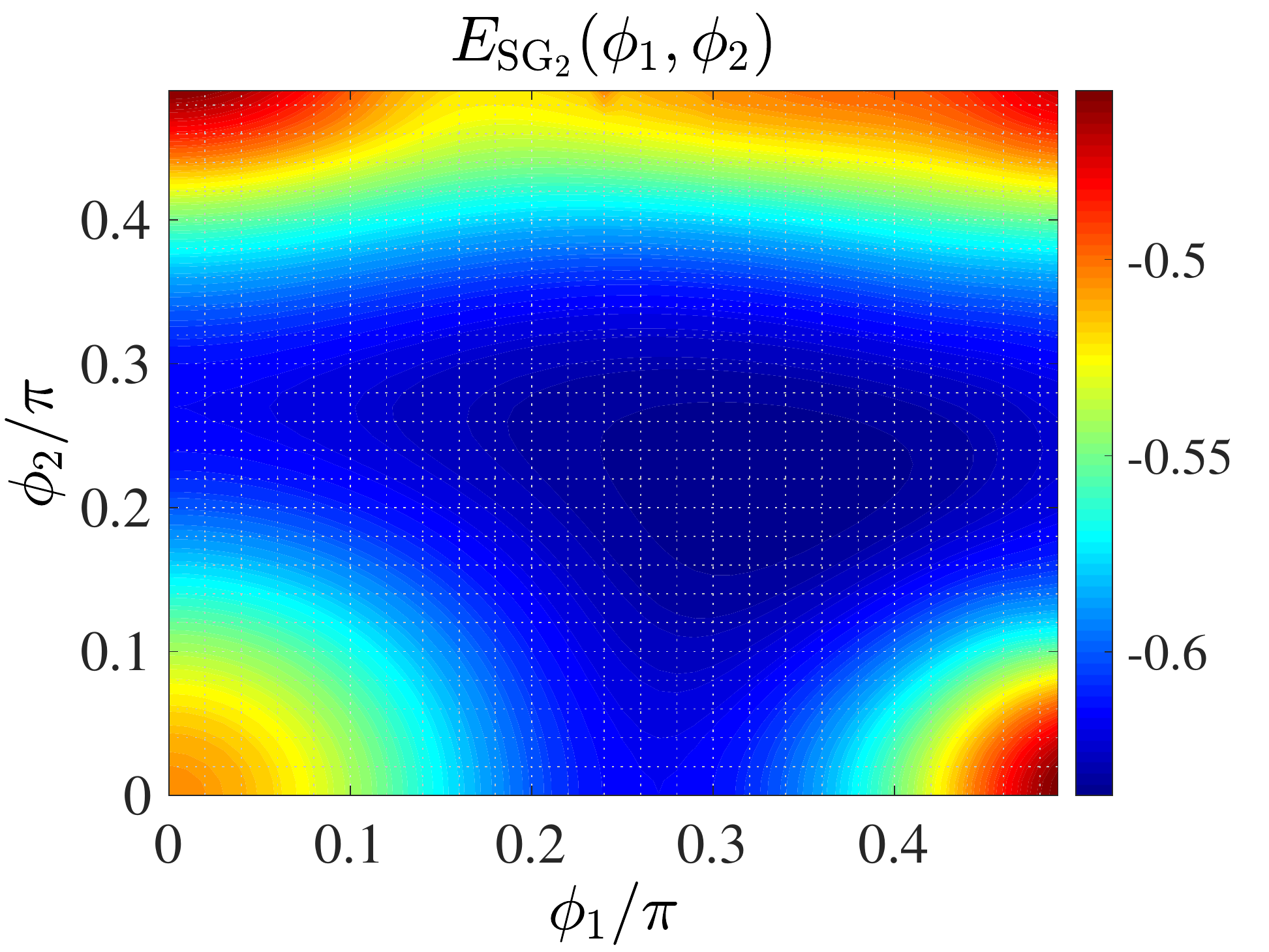}
	\caption{ The variational energy profile of the second order string gas ansatz $|\psi_{{\rm SG}_2}(\phi_1,\phi_2)\rangle$. The lowest energy is obtained at $(\phi_1,\phi_2) = (0.33\pi, 0.22\pi)$. }
	\label{fig:sg2_energy}
\end{figure}

As explained in the main text, the second order string gas ansatz has two independent variational parameters $(\phi_1, \phi_2)$, i.e., $|\psi_{{\rm SG}_2}(\phi_1,\phi_2)\rangle$. Figure\,\ref{fig:sg2_energy} presents the energy profile as a function of $(\phi_1, \phi_2)$, and the lowest energy $E_{{\rm SG}_2} = -0.6366$ is obtained at  $(\phi_1,\phi_2) = (0.33\pi, 0.22\pi)$.

\subsection{ Imaginary time evolution flow of physical quantities}

\begin{figure}[!h]
	\includegraphics[width=0.99\textwidth]{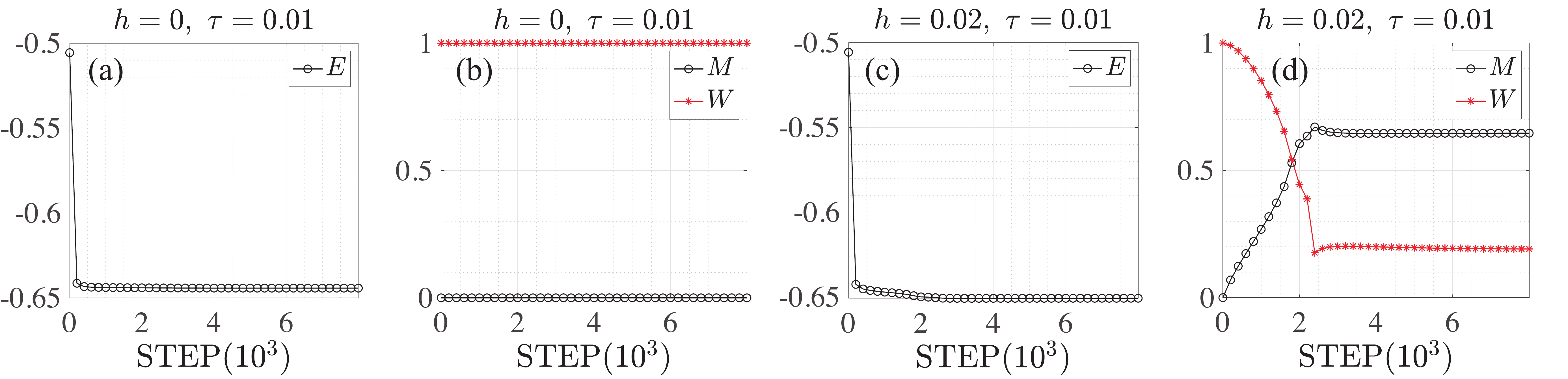}
	\caption{ Plots of the imaginary time evolution flow of (a), (c) the variational energies and (b), (d) the magnetization and flux expectation value at $h=0$ and $h=0.02$, respectively. The initial state is the loop gas state, and the bond dimension is $D=8$ and the imaginary time step is $\tau = 0.01$. }
	\label{fig:e_m_flow}
\end{figure}

The imaginary time evolution\,(ITE) starting from the loop gas state is stable and effective in the presence of weak magnetic field. As illustrative examples, we present the ITE flow of the variational energy, magnetization and flux expectation value as a function of the ITE step in Fig.\,\ref{fig:e_m_flow} at $h=0$ and $h=0.02$. As one can see, the magnetization and flux expectation value remain unity and zero, respectively, throughout the ITE flow at zero field. On the other hand, at $h=0.02$, the ITE flow exhibits transition-like behavior at a certain ITE step, at which the ansatz seems becoming the polarized state.

\subsection{ Different flux sector }

\begin{figure}[!h]
	\includegraphics[width=0.99\textwidth]{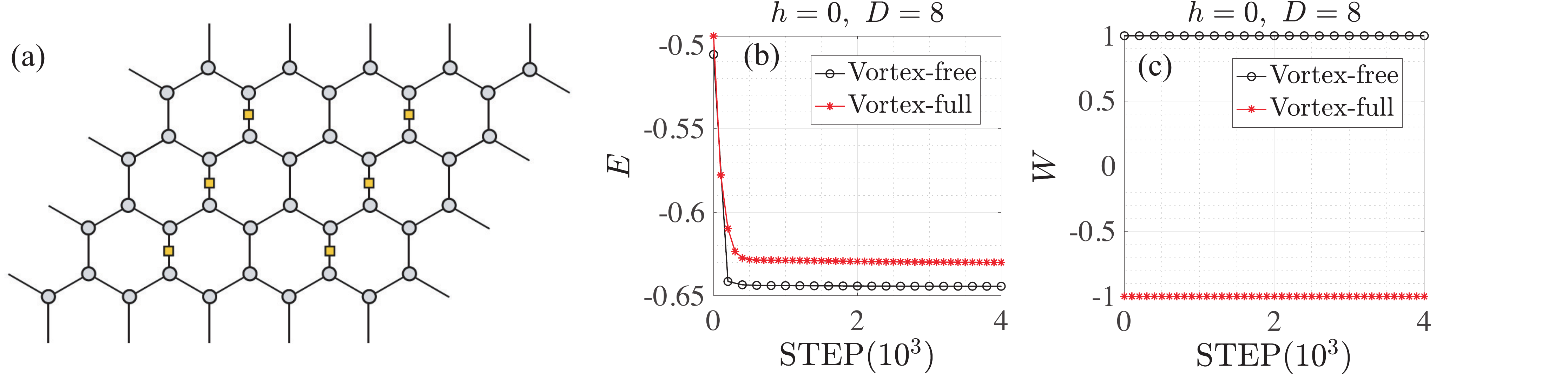}
	\caption{ (a) Schematic figure of the vortex-full TN state where the yellow square stands for the non-trivial element of $Z_2$ invariant gauge group\,[see Sec.\,\ref{sec:IGG}]. (b) The ITE flow of the variational energies in the vortex-free and vortex-full sectors with $D=8$ at zero field. }
	\label{fig:vortex_full}
\end{figure}

The LG representation allows us to access the other flux sector. For instance, an ansatz in the vortex-full sector, in which all plaquetts are occupied by the vortex $W_p=-1$, can be obtained by inserting the non-trivial element of the invariant gauge group as illustrated in Fig.\,\ref{fig:vortex_full}\,(a). Then, one can perform the ITE in a given flux sector as presented in Fig.\,\ref{fig:vortex_full}\,(b) and (c) for the vortex-free and vortex-full sectors. The ansatz in the vortex-free sector has significantly lower energy than that of the vortex-full state. We have also checked the vortex-crystal sectors and found that all those sectors are unstable compared to the vortex-free one.

\subsection{ Magnetic field effect }
\begin{figure}[!h]
	\includegraphics[width=0.99\textwidth]{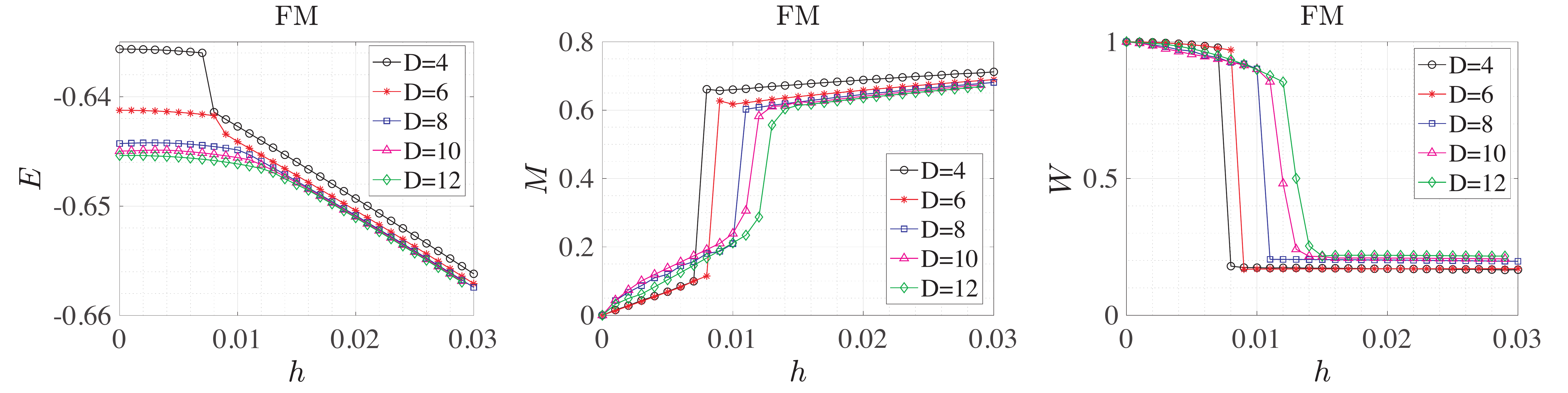}
	\caption{ Plots of the bond dimension $D$-dependence of the energy\,(left), magnetization\,(center) and flux expectation value\,(right) as a function of the field strength $h$.}
	\label{fig:d_dependence}
\end{figure}

\begin{figure}[!h]
	\includegraphics[width=0.7\textwidth]{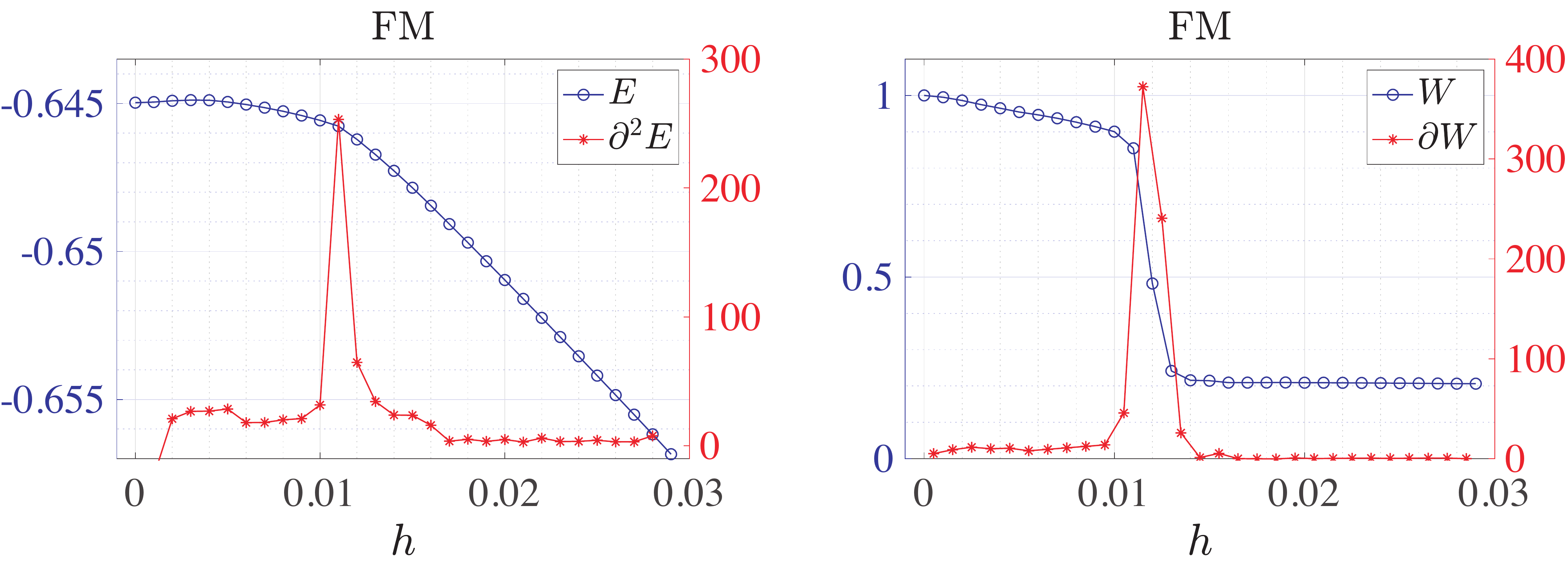}
	\caption{ Plots of (a) the variational energy and its second derivative and (b) the flux expectation value and its first derivative as a function of the field strength $h$. Here, the $D=10$ ansatze are used.}
	\label{fig:fm_field}
\end{figure}

As demonstrated in the main text, the ferromagnetic model exhibits a phase transition in the presence of the (111)-direction magnetic field. Figure\,\ref{fig:d_dependence} presents the bond dimension $D$-dependence of the energy, magnetization and flux expectation value as a function of the field strength $h$. Interestingly, when $D$ is small\,($D<10$), the transition is discontinuous, i.e., the first order transition, while it turns out to be the continuous one with $D \geq 10$. However, the critical field is still not converged up to $D=12$, so that we expect the true critical field is slightly larger than $h_c^{D=12}\approx 0.13$. We also present the energy and its second derivative, and the flux expectation value and its first derivative in Fig.\,\ref{fig:fm_field}.

\end{document}